\documentclass[prd]{revtex4}

\usepackage{amsmath,amssymb}
\usepackage{graphicx}
\usepackage{pst-plot}
\usepackage{mathrsfs}
\usepackage{mathtools}
\usepackage{pst-func}

\DeclareMathAlphabet{\mathcal}{OMS}{cmsy}{m}{n}

\usepackage{titlesec}

\titlespacing\subsection{0pt}{12pt plus 2pt minus 2pt}{4pt plus 2pt minus 2pt}

\def\dbl{\hbox{${1\hskip -2.4pt{\rm l}}$}}

\begin{document}

\title{Bell's Theorem Begs the Question}

\author{Joy Christian}

\email{jjc@bu.edu}

\affiliation{Einstein Centre for Local-Realistic Physics, Oxford OX2 6LB, United Kingdom}

\begin{abstract}
I demonstrate that Bell's theorem is based on circular reasoning and thus a fundamentally flawed argument. It unjustifiably assumes the additivity of expectation values for dispersion-free states of contextual hidden variable theories for non-commuting observables involved in Bell-test experiments, which is tautologous to assuming the bounds of $\pm2$ on the Bell-CHSH sum of expectation values. Its premisses thus assume in a different guise the bounds of $\pm2\,$ it sets out to prove. Once this oversight is ameliorated from Bell's argument by identifying the impediment that leads to it and local realism is implemented correctly, the bounds on the Bell-CHSH sum of expectation values work out to be ${\pm2\sqrt{2}}$ instead of ${\pm2}$, thereby mitigating the conclusion of Bell's theorem. Consequently, what is ruled out by any of the Bell-test experiments is not local realism but the linear additivity of expectation values,\break which does not hold for non-commuting observables in any hidden variable theories to begin with. I also identify similar oversight in the GHZ variant of Bell's theorem, invalidating its claim of having found an inconsistency in the premisses of the argument by EPR for completing quantum mechanics. Conceptually, the oversight in both Bell's theorem and its GHZ variant traces back to the oversight in\break von~Neumann's theorem against hidden variable theories identified by Grete Hermann in the 1930s. 
\end{abstract}

\maketitle

KEYWORDS: Bell's theorem, local realism, Bell-CHSH inequalities, quantum correlations, Bell-test experiments

\parskip 6pt

\parindent 12pt

\baselineskip 12.47pt

\subsection{Introduction} \label{Sec-A}

Bell's theorem \cite{Bell-1964} is an impossibility argument (or ``proof'') that claims that no locally causal and realistic hidden variable theory envisaged by Einstein \cite{Einstein} that could ``complete'' quantum theory can reproduce all of the predictions of quantum theory. But some such claims of impossibility in physics are known to harbor unjustified assumptions. In this paper, I show that Bell’s theorem against locally causal hidden variable theories is no exception. It is no different, in this respect, from von Neumann’s theorem against all hidden variable theories \cite{vonNeumann}, or the Coleman-Mandula theorem overlooking the possibilities of supersymmetry \cite{Coleman}. The implicit and unjustified assumptions underlying the latter two theorems seemed so innocuous to many that they escaped notice for decades. By contrast, Bell's theorem has faced skepticism and challenges by many from its very inception (cf. footnote~1 in \cite{IEEE-1}), including by me \cite{IEEE-1,Disproof, Christian,IJTP,Oversight,RSOS,IEEE-2,IEEE-3,IEEE-4,Local,Symmetric,RSOS-Reply}, because it depends on a number of questionable implicit and explicit physical assumptions that are not difficult to recognize \cite{RSOS,RSOS-Reply}.~In what follows, I bring out one such assumption and demonstrate that Bell's theorem is based on a circular argument \cite{Oversight}. It unjustifiably assumes the additivity of expectation values for dispersion-free states of hidden variable theories for non-commuting observables involved in the Bell-test experiments \cite{Clauser}, which is tautologous to assuming the bounds of $\pm2$ on the Bell-CHSH sum of expectation values \cite{CHSH}. Its premisses thus assume in a different guise what it sets out to prove. Once this oversight is ameliorated from Bell's argument, the local-realistic bounds on the Bell-CHSH\break sum of expectation values work out to be ${\pm2\sqrt{2}}$ instead of ${\pm2}$, thereby mitigating the conclusion of Bell's theorem. As a result, what is ruled out by the Bell-test experiments is not local realism but the additivity of expectation values, which does not hold for non-commuting observables in dispersion-free states of hidden variable theories to begin with.

This flaw, and its multiplicative analog, invalidates not only Bell's theorem \cite{Bell-1964} but also its variants \cite{CH,GHZ,Hardy}. Among its variants that rely on inequalities similar to the Bell-CHSH inequalities \cite{Bell-1964,CHSH}, such as that by Clauser and Horne (CH) \cite{CH}, the demonstration of the flaw presented below for Bell's original theorem goes through with little or no amendments. On the other hand, in the variant of Bell's theorem by Greenberger, Horne, and Zeilinger (GHZ) \cite{GHZ}, which does not involve inequalities but alleges an inconsistency in the premisses of the program set out by Einstein, Podolsky, and Rosen (EPR) \cite{EPR} for local-realistically completing quantum mechanics, the flaw takes a somewhat different form, as I explain below in Section~\ref{GHZ-flaw}. I demonstrate that the GHZ claim of inconsistency in the premisses of the EPR argument implicitly depends on multiplicative expectation functions and eigenvalues for non-commuting observables involved in their thought experiment, and is therefore just as invalid as Bell's original theorem that implicitly depends on linear additivity of expectation functions and eigenvalues. There is also a variant of Bell's theorem by Hardy that does not involve inequalities and falls halfway between Bell's original theorem and its variant by GHZ \cite{Hardy}. But in Section~\ref{Hardy-flaw} below I demonstrate that, contrary to its claim, Hardy's variant is not ``a proof of non-locality'' but an instance of the Kochen-Specker theorem \cite{Kochen}. It is also worth noting that, in \cite{Symmetric}, I have not only local-realistically reproduced, within a quaternionic 3-sphere model, the singlet correlations on which Bell's theorem is based, but, in \cite{RSOS,disproof-GHZ}, within the same model, I have also reproduced the correlations predicted by the GHZ states. Moreover, in \cite{disproof-GHZ} I have also reproduced all sixteen predictions of the quantum mechanical state considered by Hardy. 

\subsection{Heuristics for completing quantum mechanics} \label{Sec-B}

The goal of any hidden variable theory \cite{vonNeumann,Bell-1966,Gudder} is to reproduce the statistical predictions encoded in the quantum states $|\psi\rangle\in{\mathscr H}$ of physical systems using hypothetical dispersion-free states $|\psi,\,\lambda):= \{|\psi\rangle,\,\lambda\}\in{\mathscr H}\otimes{\mathscr L}$ that have no inherent statistical character, where the Hilbert space ${\mathscr H}$ is extended by the space ${\mathscr L}$ of hidden variables $\lambda$, which are hypothesized to ``complete'' the states of the physical systems as envisaged by Einstein \cite{Einstein}. If the values of $\lambda\in{\mathscr L}$ can be specified in advance, then the results of any measurements on a given physical system are uniquely determined.

To appreciate this, recall that expectation value of the square of any self-adjoint operator ${\Omega}\in{\mathscr H}$ in a normalized quantum mechanical state $|\psi\rangle$ and the square of the expectation value of ${\Omega}$ will not be equal to each other in general:
\begin{equation}
{\langle\psi|\,{\Omega}^2\,|\psi\rangle}\not={\left\langle\psi\left|\,{\Omega}\,\right|\psi\right\rangle}^2.
\end{equation}
This gives rise to inherent statistical uncertainty in the value of $\Omega$, indicating that the state $|\psi\rangle$ is not dispersion-free:
\begin{equation}
\Delta\Omega=\sqrt{\langle\psi|\{\,\Omega-\langle\psi|\,\Omega\,|\psi\rangle\dbl\}^2\,|\psi\rangle}\not=0. \label{disp}
\end{equation}
By contrast, in a normalized dispersion-free state $|\psi,\,\lambda)$ of hidden variable theories formalized by von Neumann \cite{vonNeumann}, the expectation value of ${\Omega}$, {\it by hypothesis}, is equal to one of its eigenvalues ${\omega}({\lambda})$, determined by the hidden variables$\;\lambda$,
\begin{equation}
\left(\,\psi,\,\lambda\,|\,\Omega\,|\,\psi,\,\lambda\,\right) = {\omega}({\lambda}) \;\Longleftrightarrow\; \Omega\,|\,\psi,\,\lambda) = {\omega}({\lambda})\,|\,\psi,\,\lambda), \label{hidres}
\end{equation}
so that a measurement of $\Omega$ in the state $\left|\,\psi,\,\lambda\,\right)$ would yield the result ${\omega({\lambda})}$ with certainty. How this can be accomplished in a dynamical theory of measurement process remains an open question \cite{Bell-1966}. But accepting the hypothesis (\ref{hidres}) implies
\begin{equation}
(\psi,\,\lambda\,|\,\Omega^2\,|\,\psi,\,\lambda) = (\psi,\,\lambda\,|\,\Omega\,|\,\psi,\,\lambda)^2.
\end{equation}
Consequently, unlike in a quantum sate $|\psi\rangle$, in a dispersion-free state $|\psi,\,\lambda)$ observables $\Omega$ have no inherent uncertainty:
\begin{equation}
\Delta\Omega=\sqrt{(\,\psi,\,\lambda\,|\,\{\,\Omega-\left(\,\psi,\,\lambda\,|\,\Omega\,|\,\psi,\,\lambda\,\right)\dbl\}^2\,|\,\psi,\,\lambda)}=0.
\end{equation}
The expectation value of $\Omega$ in the quantum state $|\psi\rangle$ can then be recovered by integrating over the hidden variables$\;\lambda$:
\begin{equation}
\left\langle\,\psi\,|\,\Omega\,|\,\psi\,\right\rangle \,=\int_{\mathscr L}
\left(\,\psi,\,\lambda\,|\,\Omega\,|\,\psi,\,\lambda\,\right)\,p(\lambda)\,d\lambda \,=\int_{\mathscr L}{\omega}({\lambda})\;p(\lambda)\,d\lambda\,, \label{77}
\end{equation}
where ${p(\lambda)}$ denotes the normalized probability distribution over the space ${\mathscr L}$ of thus hypothesized hidden variables. The quantum mechanical dispersion (\ref{disp}) in the measured value of the observer $\Omega$ can thus be interpreted as due to the distribution $p(\lambda)$ in the values of the hidden variables $\lambda$ over the statistical ensemble of the physical systems measured. Moreover, the Born rule can also be recovered using this prescription. If the system is in a quantum state $|\psi\rangle$ and the observable $\Omega$ satisfies the eigenvalue equation $\Omega\,|\,\omega\rangle=\omega\,|\,\omega\rangle$, then, using (\ref{77}) and the eigenvalues $\pi=1$ and $0$ of the corresponding projection operator $|\,\omega\rangle\langle\omega\,|$, the probability of observing the eigenvalue $\omega$ of $\Omega$ can be recovered as
\begin{equation}
P(\Omega\rightarrow\omega\,|\,|\psi\rangle)=|\langle\omega\,|\,\psi\,\rangle|^2=\left\langle\,\psi\,|\,\omega\rangle\langle\omega\,|\,\psi\,\right\rangle \,=\int_{\mathscr L}
\left(\,\psi,\,\lambda\,|(|\,\omega\rangle\langle\omega\,|)|\,\psi,\,\lambda\,\right)\,p(\lambda)\,d\lambda \,=\int_{\mathscr L}\pi(\lambda)\;p(\lambda)\,d\lambda\,. \label{88}
\end{equation}
The probabilities predicted by the Born rule can thus be interpreted as arising from the statistical distribution of $\lambda$. 

As it stands, prescription (\ref{77}) amounts to assignment of unique eigenvalues ${\omega}({\lambda})$ to {\it all} observables $\Omega$ {\it simultaneously}, regardless of whether they are actually measured. In other words, according to (\ref{77}) every physical quantity of a given system represented by $\Omega$ would possess a unique preexisting value, irrespective of any measurements being performed. The prescription (\ref{77}) thus mathematically encodes Einstein's conception of realism. In \cite{Einstein}, Einstein explained his point of view in terms of the position and momentum of a free particle as follows: “The (free) particle really has a definite position and a definite momentum, even if they cannot both be ascertained by measurement in the same individual case [as in quantum mechanics]. According to this point of view, the $\psi$-function represents an incomplete description of the real state of affairs.” The acceptance of this point of view --- Einstein continues --- “would lead to an attempt to\break obtain a complete description of the real state of affairs {\it as well as the incomplete one} [my emphasis], and to discover physical laws for such a description.” Accordingly, the left-hand side of the first equality in (\ref{77}) provides the incomplete description of the system and its right-hand side provides the complete one, with all possible statements one can make about the system encoded in the expectation values of the observables being measured in the state of the system \cite{vonNeumann}. If $\Omega_1$ and $\Omega_2$ are two non-commuting observables, then the uncertainty relation between them can also be interpreted~as
\begin{equation}
\Delta\Omega_1 \, \Delta\Omega_2 \,\geqslant\, \frac{1}{2}\left| \left\langle\,\psi\,|\,\left[\Omega_1,\,\Omega_2\right]\,|\,\psi\,\right\rangle \right| =
\frac{1}{2}\left| \int_{\mathscr L}
\left(\,\psi,\,\lambda\,|\,\left[\Omega_1,\,\Omega_2\right]\,|\,\psi,\,\lambda\,\right)\,p(\lambda)\,d\lambda\, \right| = \frac{1}{2}\left|\int_{\mathscr L}{\omega_{1,2}}({\lambda})\;p(\lambda)\,d\lambda\,\right| \label{7in}
\end{equation}
in terms of the probability distribution ${p(\lambda)}$ in the values of the hidden variables $\lambda$, where the first inequality is the one established by Robertson \cite{Robertson}, and $i\,\omega_{1,2}(\lambda)$ is a purely imaginary eigenvalue of the skew-Hermitian operator $\left[\Omega_1,\,\Omega_2\right]$. Similarly, using the kinematical equivalence seen in (\ref{77}), the dynamical equivalence between the quantum mechanical description and Einstein's ``complete'' description can also be established, as demonstrated in the Appendix~\ref{D} below:
\begin{equation}
\left[\frac{d\,}{dt}\left\langle\,\psi\,|\,\Omega\,|\,\psi\,\right\rangle = \frac{1}{i\hbar}\left\langle\,\psi\,|\,\left[\,\Omega,\,H\,\right]\,|\,\psi\,\right\rangle + \langle\,\psi\,|\,\frac{\partial\Omega}{\partial t}\,|\,\psi\,\rangle\right] = \int_{\mathscr L}\left[\frac{d\;}{dt}\,\omega(\lambda) = \{\omega(\lambda),\,{\cal H}(\lambda)\} +\frac{\partial\,\omega(\lambda)}{\partial t}\right] \,p(\lambda)\,d\lambda\,, \label{Eh}
\end{equation}
where $H$ is a Hamiltonian operator in ${\mathscr H}$ and ${\cal H}$ is a classical Hamiltonian function in the corresponding phase space. If particular values of $\lambda$ are precisely known with $p(\lambda)=1$, then the right-hand side of (\ref{Eh}) would reduce to the classical Hamiltonian equations of motion.~Otherwise, Ehrenfest's equation in quantum mechanics on the left-hand side of (\ref{Eh}) can be understood as an ensemble average of classical dynamics over the probability distribution $p(\lambda)$ of $\lambda$.~Thus, once the hypothesis (\ref{hidres}) regarding the dispersion-free states $|\psi,\,\lambda)$ is accepted, each probabilistic statement about the quantum system, (\ref{77}), (\ref{88}), (\ref{7in}), and (\ref{Eh}), can be traced back to the incompleteness of our knowledge about the system.

In Section 2 of \cite{Bell-1966}, Bell works out an instructive example to illustrate how the prescription (\ref{77}) works for a system of two-dimensional Hilbert space. It fails, however, for Hilbert spaces of dimensions greater than two, because in higher dimensions degeneracies prevent simultaneous assignments of unique eigenvalues to all observables in dispersion-free states $\left|\,\psi,\,\lambda\,\right)$ dictated by the ansatz (\ref{hidres}), giving contradictory values for the same physical quantities.~This was proved independently by Bell \cite{Bell-1966}, Kochen and Specker \cite{Kochen}, and Belinfante \cite{Belinfante}, as a corollary to Gleason's theorem \cite{Gleason,Shimony}.

These proofs -- known as the Kochen-Specker theorem -- do not exclude contextual hidden variable theories in which the complete state $|\,\psi,\,\lambda)$ of a system assigns unique values to physical quantities only {\it relative} to experimental contexts \cite{Gudder,Shimony}. If we denote the observables as $\Omega(c)$ with $c$ being the environmental contexts of their measurements, then the\break non-contextual prescription (\ref{77}) can be easily modified to accommodate contextual hidden variable theories as follows:
\begin{equation}
\left\langle\,\psi\,|\,\Omega(c)\,|\,\psi\,\right\rangle \,=\int_{\mathscr L}
\left(\,\psi,\,\lambda\,|\,\Omega(c)\,|\,\psi,\,\lambda\,\right)\,p(\lambda)\,d\lambda \,=
\int_{\mathscr L}{\omega}(c,\,\lambda)\;p(\lambda)\,d\lambda\,. \label{99}
\end{equation}
Each observable $\Omega(c)$ is still assigned a unique eigenvalue ${\omega}(c,\,\lambda)$, but now determined cooperatively by the complete state $|\,\psi,\,\lambda)$ of the system and the state $c$ of its environmental contexts. Consequently, even though some of its features are no longer intrinsic to the system, contextual hidden variable theories do not have the inherent statistical character of quantum mechanics, because outcome of an experiment is a cooperative effect just as it is in classical physics \cite{Shimony}.\break Therefore, such theories interpret quantum entanglement at the level of the complete state $|\,\psi,\,\lambda)$ only epistemically.

\subsubsection{Expectation function $\left(\,\psi,\,\lambda\,|\,\Omega(c)\,|\,\psi,\,\lambda\,\right) $ for non-commuting observables cannot be linear}

For our purposes here, it is also important to recall that in the Hilbert space formulation of quantum mechanics \cite{vonNeumann} the correspondence between observables and Hermitian operators is one-to-one. Moreover, a sum $\widetilde{\Omega}(\tilde{c})=\sum_{i=1}^n\Omega_i(c_i)$ of several observables such as $\Omega_1(c_1),\,\Omega_2(c_2),\,\Omega_3(c_3),\dots,\,\Omega_n(c_n)$ is also an observable representing a physical quantity, and consequently the sum of the expectation values of $\Omega_i(c_i)$ is the expectation value of the summed operator $\widetilde{\Omega}(\tilde{c})$,
\begin{equation}
 \sum_{i=1}^n\left\langle\,\psi\,|\,\Omega_i(c_i)\,|\,\psi\,\right\rangle=\langle\,\psi\,|\left[\sum_{i=1}^n\Omega_i(c_i)\right]|\,\psi\,\rangle, \label{sum}
\end{equation}
regardless of whether the observables are simultaneously measurable or mutually commutative \cite{Bell-1966}. The question then is, since within any contextual hidden variable theory characterized by (\ref{99}) all of the observables $\Omega_i(c_i)$ and their sum\break $\widetilde{\Omega}(\tilde{c})$ are assigned unique eigenvalues $\omega_i(c_i,\,\lambda)$ and $\widetilde{\omega}(\tilde{c},\,\lambda)$, respectively, would these eigenvalues satisfy the equality
\begin{equation}
\sum_{i=1}^n\left[\int_{\mathscr L}{\omega_i}(c_i,\,\lambda)\;p(\lambda)\,d\lambda\right]\,\stackrel{?}{=}\int_{\mathscr L}\left[\sum_{i=1}^n {\omega_i}(c_i,\,\lambda)\right]p(\lambda)\,d\lambda \label{fol9}
\end{equation}
in dispersion-free states $|\,\psi,\,\lambda)$ of physical systems in analogy with the linear quantum mechanical relation (\ref{sum}) above?\break The answer is: Not in general, because the eigenvalue ${\widetilde{\omega}(\tilde{c},\,\lambda)}$ of the summed operator $\widetilde{\Omega}(\tilde{c})$ is not equal to the sum $\sum_{i=1}^n\omega_i(c_i,\,\lambda)$ of eigenvalues $\omega_i(c_i,\,\lambda)$ for given $\lambda$, unless the constituent observables $\Omega_i(c_i)$ are mutually commutative. In other words, ${\widetilde{\omega}(\tilde{c},\,\lambda)}\not=\sum_{i=1}^n\omega_i(c_i,\,\lambda)$ in general, and therefore the correct counterpart of relation (\ref{sum}) is not~(\ref{fol9})~but
\begin{equation}
\sum_{i=1}^n\left[\int_{\mathscr L}{\omega_i}(c_i,\,\lambda)\;p(\lambda)\,d\lambda\right]\,=\int_{\mathscr L}\widetilde{\omega}(\tilde{c},\,\lambda)\;p(\lambda)\,d\lambda\,. \label{notfl9}
\end{equation}
As Bell points out in Section 3 of \cite{Bell-1966}, the linear relation (\ref{sum}) is an unusual property of quantum mechanical states $|\psi\rangle$.~There is no reason to demand it {\it individually} of the dispersion-free states $|\,\psi,\,\lambda)$, whose function is to reproduce the\break measurable features of quantum systems only when averaged over, as in (\ref{99}). There is no reason why the value of $\Omega(c)$\break should not be determined by some {\it nonlinear} function $\left(\,\psi,\,\lambda\,|\,\Omega(c)\,|\,\psi,\,\lambda\,\right)$. I will come back to this issue in Section~\ref{Sec-E}.

In \cite{Bell-1966}, Bell explains this non-linearity using spin components of a spin-$\frac{1}{2}$ particle. If we measure the $\sigma_x$ component of the spin with a Stern-Gerlach magnet suitably oriented in $\mathrm{I\!R}^3$, then it would yield an eigenvalue $s_x$ of $\sigma_x$ as a result. However, if we measure the $\sigma_y$ component of the spin, then that would require a different orientation of the magnet in $\mathrm{I\!R}^3$, and would give a different eigenvalue, $s_y$ of $\sigma_y$, as a result. Moreover, since $[\sigma_x,\,\sigma_y]\not=0$, a measurement of the sum of the $x$- and $y$-components of the spin, $\sigma_x+\sigma_y$, would again require a very different orientation of the magnet in $\mathrm{I\!R}^3$. Therefore, the result obtained as an eigenvalue of the summed operators $\sigma_x+\sigma_y$ will not be the sum $s_x+s_y$ of an eigenvalue of the operator $\sigma_x$ added linearly to an eigenvalue of the operator $\sigma_y$. Indeed, the eigenvalues of $\sigma_x$ and $\sigma_y$ are both $\pm1$, while the eigenvalues of $\sigma_x+\sigma_y$ are $\pm\sqrt{2}$, so a linear relation cannot hold. As Bell points out\break in \cite{Bell-1966}, the additivity of expectation values, $\langle\,\psi\,|\,\sigma_x\,|\,\psi\,\rangle+\langle\,\psi\,|\,\sigma_y\,|\,\psi\,\rangle=\langle\,\psi\,|\,\sigma_x+\,\sigma_y\,|\,\psi\,\rangle$, is a rather unusual property of the quantum states $|\psi\rangle$. The linearity of it is effectuated in quantum mechanics by promoting observable quantities to self-adjoint operators \cite{Grete}. It does not hold for the dispersion-free states $|\psi,\,\lambda)$ of hidden variable theories in general because the eigenvalues of non-commuting observables such as $\sigma_x$ and $\sigma_y$ do not add linearly, as we noted above. Consequently, the additivity relation (\ref{sum}) that holds for quantum states would not hold for the dispersion-free states.

\subsection{Special case of the singlet state and EPR-Bohm observables} \label{Sec-C}

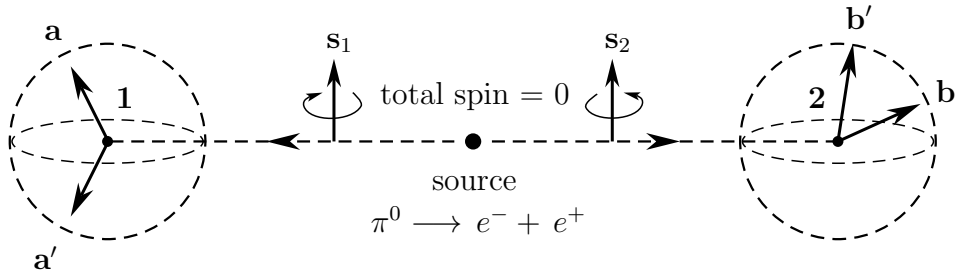
\begin{figure*}[t]
\hrule
\scalebox{1}{
\begin{pspicture}(1.0,-2.3)(4.0,2.4)

\psline[linewidth=0.1mm,dotsize=3pt 4]{*-}(-2.51,0)(-2.5,0)

\psline[linewidth=0.1mm,dotsize=3pt 4]{*-}(7.2,0)(7.15,0)

\psline[linewidth=0.4mm,arrowinset=0.3,arrowsize=3pt 3,arrowlength=2]{->}(-2.5,0)(-3,1)

\psline[linewidth=0.4mm,arrowinset=0.3,arrowsize=3pt 3,arrowlength=2]{->}(-2.5,0)(-3,-1)

\psline[linewidth=0.4mm,arrowinset=0.3,arrowsize=3pt 3,arrowlength=2]{->}(7.2,0)(8.3,0.5)

\psline[linewidth=0.4mm,arrowinset=0.3,arrowsize=3pt 3,arrowlength=2]{->}(7.2,0)(7.4,1.3)

\psline[linewidth=0.4mm,arrowinset=0.3,arrowsize=2pt 3,arrowlength=2]{->}(4.2,0)(4.2,1.1)

\psline[linewidth=0.4mm,arrowinset=0.3,arrowsize=2pt 3,arrowlength=2]{->}(0.5,0)(0.5,1.1)

\pscurve[linewidth=0.2mm,arrowinset=0.2,arrowsize=2pt 2,arrowlength=2]{->}(4.0,0.63)(3.85,0.45)(4.6,0.5)(4.35,0.65)

\put(4.1,1.25){{\large ${{\bf s}_2}$}}

\pscurve[linewidth=0.2mm,arrowinset=0.2,arrowsize=2pt 2,arrowlength=2]{<-}(0.35,0.65)(0.1,0.47)(0.86,0.47)(0.75,0.65)

\put(0.4,1.25){{\large ${{\bf s}_1}$}}

\put(-2.4,+0.45){{\large ${\bf 1}$}}

\put(6.8,+0.45){{\large ${\bf 2}$}}

\put(-3.35,1.35){{\large ${\bf a}$}}

\put(-3.5,-1.7){{\large ${\bf a'}$}}

\put(8.5,0.52){{\large ${\bf b}$}}

\put(7.3,1.5){{\large ${\bf b'}$}}

\put(1.8,-0.65){\large source}

\put(0.99,-1.2){\large ${\pi^0\longrightarrow\,e^{-}+\,e^{+}\,}$}

\put(1.11,0.5){\large total spin = 0}

\psline[linewidth=0.3mm,linestyle=dashed](-2.47,0)(2.1,0)

\psline[linewidth=0.4mm,arrowinset=0.3,arrowsize=3pt 3,arrowlength=2]{->}(-0.3,0)(-0.4,0)

\psline[linewidth=0.3mm,linestyle=dashed](2.6,0)(7.2,0)

\psline[linewidth=0.4mm,arrowinset=0.3,arrowsize=3pt 3,arrowlength=2]{->}(5.0,0)(5.1,0)

\psline[linewidth=0.1mm,dotsize=5pt 4]{*-}(2.35,0)(2.4,0)

\pscircle[linewidth=0.3mm,linestyle=dashed](7.2,0){1.3}

\psellipse[linewidth=0.2mm,linestyle=dashed](7.2,0)(1.28,0.3)

\pscircle[linewidth=0.3mm,linestyle=dashed](-2.51,0){1.3}

\psellipse[linewidth=0.2mm,linestyle=dashed](-2.51,0)(1.28,0.3)

\end{pspicture}}
\hrule
\caption{In an EPR-Bohm-type experiment, a spin-less fermion -- such as a neutral pion -- is assumed to decay from a source into an electron-positron pair, as depicted.~Then, measurements of the spin components of each separated fermion are performed at space-like separated observation stations ${\mathbf{1}}$ and ${\mathbf{2}}$, obtaining binary results $\mathscr{A}=\pm1$ and $\mathscr{B}=\pm1$ along directions ${\mathbf a}$ and ${\mathbf b}$. The conservation of spin momentum dictates that the total spin of the system remains zero during its free evolution. After Ref.~\cite{IEEE-1}.}
\label{Fig-1}
\smallskip
\smallskip
\hrule
\end{figure*}

Now, the proof of Bell's famous theorem \cite{Bell-1964} is based on Bohm's spin version of the EPR's thought experiment \cite{Bohm-1951}, which involves an entangled pair of spin-$\frac{1}{2}$ particles emerging from a source and moving freely in opposite directions, with particles ${1}$ and ${2}$ subject, respectively, to spin measurements along independently chosen unit directions ${\bf a}$ and ${\bf b}$ by Alice and Bob, who are stationed at a spacelike separated distance from each other (see Fig.~\ref{Fig-1}).~If initially the pair has vanishing total spin, then the quantum mechanical state of the system is described by the entangled singlet~state
\begin{equation}
|\Psi\rangle=\frac{1}{\sqrt{2}}\Bigl\{|{\bf k},\,+\rangle_1\otimes
|{\bf k},\,-\rangle_2\,-\,|{\bf k},\,-\rangle_1\otimes|{\bf k},\,+\rangle_2\Bigr\},
\label{single}
\end{equation}
where ${\bf k}$ is a unit vector in arbitrary direction in ${\mathrm{I\!R}^3}$ and the eigenvalue equation 
\begin{equation}
{\boldsymbol\sigma}\cdot{\bf k}\,|{\bf k},\,\pm\rangle\,=\,
\pm\,|{\bf k},\,\pm\rangle \label{spin}
\end{equation}
defines quantum mechanical eigenstates in which the two fermions have spins ``up'' or ``down'' in the units of ${\hbar=2}$, with ${\boldsymbol\sigma}$ being the Pauli spin ``vector'' ${({\sigma_x},\,{\sigma_y},\,{\sigma_z})}$. Once the state (\ref{single}) is prepared, the observable $\Omega(c)$ of interest is
\begin{equation}
\Omega(c)= {\boldsymbol\sigma}_1\cdot{\bf a}\,\otimes\,{\boldsymbol\sigma}_2\cdot{\bf b}\,, \label{obs}
\end{equation}
whose possible eigenvalues, written in terms of the dispersion-free state $|\Psi,\,\lambda)$ instead of the quantum state (\ref{single}), are
\begin{equation}
\omega(c,\,\lambda) = {\mathscr A}{\mathscr B}({\bf a},\,{\bf b},\lambda)=\pm1, \label{eig}
\end{equation}
where ${\mathscr A}=\pm1$ and ${\mathscr B}=\pm1$ are the results of spin measurements made jointly by Alice and Bob along their randomly chosen detector directions ${\bf a}$ and ${\bf b}$. In the singlet state (\ref{single}), the joint observable (\ref{obs}) predicts sinusoidal correlations $\langle\Psi|{\boldsymbol\sigma}_1\cdot{\bf a}\,\otimes\,{\boldsymbol\sigma}_2\cdot{\bf b}|\Psi\rangle=-{\bf a}\cdot{\bf b}$ between the values of the spins observed about the freely chosen contexts ${\bf a}$ and ${\bf b}$ \cite{Christian}. 

For {\it locally} contextual hidden variable theories there is a further requirement that the results of local measurements must be describable by functions that respect local causality, as first envisaged by Einstein \cite{Einstein} and later formulated mathematically by Bell \cite{Bell-1964}. It can be satisfied by requiring that the eigenvalue $\omega(c,\lambda)$ of the observable $\Omega(c)$ in (\ref{obs}) representing the joint result ${\mathscr A}{\mathscr B}({\bf a},\,{\bf b},\lambda)=\pm1$ is factorizable as $\omega(c,\lambda)=\omega_1(c_1,\lambda)\,\omega_2(c_2,\lambda)$, or in Bell's notation~as
\begin{equation}
{\mathscr A}{\mathscr B}({\bf a},\,{\bf b},\lambda)={\mathscr A}({\bf a},\lambda)\,{\mathscr B}({\bf b},\lambda), \label{fact}
\end{equation}
with the factorized functions ${\mathscr A}({\bf a},\lambda)=\pm1$ and ${\mathscr B}({\bf b},\lambda)=\pm1$ satisfying the following condition of local causality:
\begin{quote}
Apart from the hidden variables ${\lambda}$, the result ${{\mathscr A}=\pm1}$ of Alice depends {\it only} on the measurement context ${\bf a}$, chosen freely by Alice, regardless of Bob's actions \cite{Bell-1964}. And, likewise, apart from the hidden variables ${\lambda}$, the result ${{\mathscr B}=\pm1}$ of Bob depends {\it only} on the measurement context ${\bf b}$, chosen freely by Bob, regardless of Alice's actions. In particular, the function ${{\mathscr A}({\bf a},\,\lambda)}$ {\it does not} depend on ${\bf b}$ or ${\mathscr B}$ and the function ${{\mathscr B}({\bf b},\,\lambda)}$ {\it does not} depend on ${\bf a}$ or ${\mathscr A}$. Moreover, the hidden variables ${\lambda}$ do not depend on either ${\bf a}$, ${\bf b}$, ${\mathscr A}$, or ${\mathscr B}$ \cite{IEEE-2}.
\end{quote}
The expectation value ${\cal E}({\mathbf a},{\mathbf b})$ of the joint results in the dispersion-free state $|\,\psi,\,\lambda)$ should then satisfy the condition
\begin{equation}
\langle\Psi|\,{\boldsymbol\sigma}_1\cdot{\bf a}\,\otimes\,{\boldsymbol\sigma}_2\cdot{\bf b}\,|\Psi\rangle=\,{\cal E}({\mathbf a},{\mathbf b}):=\!\int_{\mathscr L} {\mathscr A}({\mathbf a},\,\lambda)\,{\mathscr B}({\mathbf b},\,\lambda)\;p(\lambda)\,d\lambda\,, \label{first}
\end{equation}
where the hidden variables ${\lambda}$ originate from a source located in the overlap of the backward light cones of Alice and Bob, and the normalized probability distribution ${p(\lambda)}$ is assumed to remain statistically independent of the contexts ${\bf a}$ and ${\bf b}$ so that $p(\lambda\,|\,{\mathbf a},{\mathbf b})=p(\lambda)$, which is a reasonable assumption. In fact, relaxing this assumption to allow $p(\lambda)$ to depend on  ${\mathbf a}$ and ${\mathbf b}$ introduces a form of non-locality, as explained by Clauser and Horne in footnote 13 of \cite{CH}.~Then, since ${\mathscr A}({\bf a},\lambda)=\pm1$ and ${\mathscr B}({\bf b},\lambda)=\pm1$, their product ${\mathscr A}({\mathbf a},\,\lambda)\,{\mathscr B}({\mathbf b},\,\lambda)=\pm1$, setting the following bounds on ${\cal E}({\mathbf a},{\mathbf b})$:
\begin{equation}
-1\leqslant\,{\cal E}({\mathbf a},{\mathbf b})\leqslant +1. \label{bounds}
\end{equation}
These bounds are respected not only by local hidden variable theories but also by quantum mechanics and experiments.

\subsection{Mathematical core of Bell's theorem} \label{Sec-D}

By contrast, at the heart of Bell's theorem is a derivation of the bounds of $\pm2$ on an {\it ad hoc} sum of the expectation values ${\cal E}({\mathbf a},{\mathbf b})$ of local results ${\mathscr A}({\bf a},\lambda)$ and ${\mathscr B}({\bf b},\lambda)$, recorded at remote observation stations by Alice and Bob, from four different sub-experiments involving measurements of non-commuting observables such as ${\boldsymbol\sigma}_1\cdot{\bf a}$ and ${\boldsymbol\sigma}_1\cdot{\bf a'}$ \cite{Bell-1964,Clauser,CHSH}:
\begin{equation}
{\cal E}({\bf a},\,{\bf b})+{\cal E}({\bf a},\,{\bf b'})+{\cal E}({\bf a'},\,{\bf b})-{\cal E}({\bf a'},\,{\bf b'})\,. \label{combi}
\end{equation}
Alice can freely choose a detector direction ${\bf a}$ or ${\bf a'}$, and likewise Bob can freely choose a detector direction ${\bf b}$ or ${\bf b'}$, to detect, at a space-like distance from each other, the spins of fermions they receive from the common source. Then, from (\ref{bounds}), we can immediately read off the upper and lower bounds on the combination (\ref{combi}) of expectation values:
\begin{equation}
-4\,\leqslant\,{\cal E}({\bf a},\,{\bf b})\,+\,{\cal E}({\bf a},\,{\bf b'})\,+\,{\cal E}({\bf a'},\,{\bf b})\,-\,{\cal E}({\bf a'},\,{\bf b'})\,\leqslant\,+4\,. \label{8-2}
\end{equation}

\subsubsection{Standard derivation of the Bell-CHSH inequalities (\ref{chsh})}

The next step in Bell's derivation of the bounds $\pm2$ instead of $\pm4$ is the assumption of the additivity of expectation values, which amounts to assuming that any sum of expectation values is equal to the expectation value of the sum:
\begin{align}
&{\cal E}({\bf a},\,{\bf b})\,+\,{\cal E}({\bf a},\,{\bf b'})\,+\,{\cal E}({\bf a'},\,{\bf b})\,-\,{\cal E}({\bf a'},\,{\bf b'}) \notag \\
&=\!\int_{\mathscr L}{\mathscr A}({\bf a},\lambda){\mathscr B}({\bf b},\lambda)\,p(\lambda)\,d\lambda+\!\!\int_{\mathscr L}{\mathscr A}({\bf a},\lambda){\mathscr B}({\bf b'},\lambda)\,p(\lambda)\,d\lambda+\!\!\!\int_{\mathscr L}\!\!{\mathscr A}({\bf a'},\lambda){\mathscr B}({\bf b},\lambda)\,p(\lambda)\,d\lambda-\!\!\int_{\mathscr L}\!\!{\mathscr A}({\bf a'},\lambda){\mathscr B}({\bf b'},\lambda)\,p(\lambda)\,d\lambda \notag \\
&=\!\int_{\mathscr L}\!\big\{\,{\mathscr A}({\bf a},\lambda)\,{\mathscr B}({\bf b},\lambda)+{\mathscr A}({\bf a},\lambda)\,{\mathscr B}({\bf b'},\lambda)+{\mathscr A}({\bf a'},\lambda)\,{\mathscr B}({\bf b},\lambda)-{\mathscr A}({\bf a'},\lambda)\,{\mathscr B}({\bf b'},\lambda)\big\}\;p(\lambda)\,d\lambda\,. \label{ladd}
\end{align}
We will have much to discuss about this seemingly harmless mathematical step employing the built-in linear additivity of integrals. I will demonstrate in Section$\,$\ref{Sec-E} that, far from being harmless, it is, in fact, {\it an unjustified assumption that harbors a profound mistake of assuming the very thesis of the theorem to be proven}, just as von Neumann's theorem did \cite{Oversight,Bell-1966,Grete,Mermin,BohmBub}. It assumes, {\it without proof}, that linear additivity of integrals leading to (\ref{ladd}) can be meaningfully applied, not only to the eigenvalues of commuting observables but also to the eigenvalues of non-commuting observables that cannot be measured simultaneously. But, as we saw in the paragraph following equation (\ref{notfl9}), this assumption is quite mistaken. However, if we overlook this mistake and accept equality (\ref{ladd}), then the bounds of $\pm2$ on the Bell-CHSH\break combination (\ref{combi}) of expectation values are not difficult to work out by rewriting the 
integrand on its right-hand~side~as
\begin{equation}
{\mathscr A}_{}({\bf a},\lambda)\,\big\{\,{\mathscr B}_{}({\bf b},\lambda)+{\mathscr B}_{}({\bf b'},\lambda)\,\big\}\,+\,{\mathscr A}_{}({\bf a'},\lambda)\,\big\{\,{\mathscr B}_{}({\bf b},\lambda)-{\mathscr B}_{}({\bf b'},\lambda)\,\big\}. \label{int}
\end{equation}
Since ${{\mathscr B}_{}({\bf b},\lambda)=\pm1}$, if ${|{\mathscr B}_{}({\bf b},\lambda)+{\mathscr B}_{}({\bf b'},\lambda)|=2}$, then ${|{\mathscr B}_{}({\bf b},\lambda)-{\mathscr B}_{}({\bf b'},\lambda)|=0}$, and vice versa. Consequently, since ${{\mathscr A}_{}({\bf a},\lambda)=\pm1}$, the integrand (\ref{int}) is bounded by $\pm2$ and the absolute value of the last integral in (\ref{ladd}) does not exceed$\;$2: 
\begin{equation}
-2\,\leqslant\int_{\mathscr L}\!\big\{\,{\mathscr A}({\bf a},\lambda)\,{\mathscr B}({\bf b},\lambda)+{\mathscr A}({\bf a},\lambda)\,{\mathscr B}({\bf b'},\lambda)+{\mathscr A}({\bf a'},\lambda)\,{\mathscr B}({\bf b},\lambda)-{\mathscr A}({\bf a'},\lambda)\,{\mathscr B}({\bf b'},\lambda)\big\}\;p(\lambda)\,d\lambda\;\leqslant\,+2\,.\label{not5}
\end{equation}
Therefore, the equality (\ref{ladd}) implies that the absolute value of the combination of expectation values is bounded by 2:
\begin{equation}
-2\,\leqslant\,{\cal E}({\bf a},\,{\bf b})\,+\,{\cal E}({\bf a},\,{\bf b'})\,+\,{\cal E}({\bf a'},\,{\bf b})\,-\,{\cal E}({\bf a'},\,{\bf b'})\,\leqslant\,+2\,. \label{chsh}
\end{equation}
But since the bounds on (\ref{combi}) predicted by quantum mechanics and observed in experiments are $\pm2\sqrt{2}$, Bell concludes that no local and realistic theory envisaged by Einstein can reproduce the statistical predictions of quantum mechanics. In particular, contextual hidden variable theories specified by (\ref{99}) that respect the factorizability (\ref{fact}) are not viable.

In many derivations of (\ref{chsh}) in the literature, factorized probabilities of observing binary measurement results are employed rather than measurement results themselves I have used in (\ref{fact}) in my derivation following Bell \cite{Bell-1964,Clauser}. But employing probabilities would only manage to obfuscate the logical flaw in Bell’s argument I intend to bring out here.

\subsubsection{Converse derivation of the additivity (\ref{ladd}) by assuming (\ref{chsh})}

Now, it is not difficult to demonstrate the {\it converse} of the above derivation in which the additivity of expectation values (\ref{ladd}) is derived by assuming the stringent bounds of $\pm2$ on the sum (\ref{combi}). Employing (\ref{first}), (\ref{combi}) can be written$\;$as
\begin{equation}
\int_{\mathscr L}{\mathscr A}({\bf a},\lambda){\mathscr B}({\bf b},\lambda)\,p(\lambda)\,d\lambda+\!\!\int_{\mathscr L}{\mathscr A}({\bf a},\lambda){\mathscr B}({\bf b'},\lambda)\,p(\lambda)\,d\lambda+\!\!\!\int_{\mathscr L}\!\!{\mathscr A}({\bf a'},\lambda){\mathscr B}({\bf b},\lambda)\,p(\lambda)\,d\lambda-\!\!\int_{\mathscr L}\!\!{\mathscr A}({\bf a'},\lambda){\mathscr B}({\bf b'},\lambda)\,p(\lambda)\,d\lambda\,. \label{23}
\end{equation}
Since each product ${\mathscr A}({\bf a},\lambda){\mathscr B}({\bf b},\lambda)$ in the above integrals is equal to $\pm1$, each of the four integrals is bounded by $\pm1$:
\begin{equation}
-1\,\leqslant\int_{\mathscr L}{\mathscr A}({\bf a},\lambda){\mathscr B}({\bf b},\lambda)\,p(\lambda)\,d\lambda\;\leqslant\,+1. \label{2333}
\end{equation}
Thus the sum of four integrals in (\ref{23}) is bounded by $\pm4$, not $\pm2$.~However, we started with (\ref{chsh}), which contends that the sum of integrals in (\ref{23}) is bounded by $\pm2$. But the only way to reduce the bounds on (\ref{23}) from $\pm4$ to $\pm2$ without compromising the independence of the results ${\mathscr A}({\bf a},\lambda){\mathscr B}({\bf b},\lambda)$, {\it etc.}, or of their averages appearing in the sum (\ref{23}), or\break violating the rules of anti-derivatives, is by equating the sum of integrals in (\ref{23}) to the following integral of the sum:
\begin{equation}
\int_{\mathscr L}\!\big\{\,{\mathscr A}({\bf a},\lambda)\,{\mathscr B}({\bf b},\lambda)+{\mathscr A}({\bf a},\lambda)\,{\mathscr B}({\bf b'},\lambda)+{\mathscr A}({\bf a'},\lambda)\,{\mathscr B}({\bf b},\lambda)-{\mathscr A}({\bf a'},\lambda)\,{\mathscr B}({\bf b'},\lambda)\big\}\;p(\lambda)\,d\lambda\,. \label{home} 
\end{equation}
To be sure, we can easily reduce the bounds on the sum in (\ref{23}) from $\pm4$ to $\pm2$, for example, by setting the last two integrals in it equal to each other, with the minus sign between them. That would reduce (\ref{23}) to a sum of only the first two integrals, which is then easily seen to be bounded by $\pm2$ using (\ref{2333}). But to achieve this, we have compromised the independence of at least the last two sub-ensemble averages, which is not permitted by the conditions adhered to in Bell's argument. Similarly, again as an example, we can set the results themselves in the last two integrals in (\ref{23}) equal to each other: ${\mathscr A}({\bf a'},\lambda)\,{\mathscr B}({\bf b},\lambda)={\mathscr A}({\bf a'},\lambda)\,{\mathscr B}({\bf b'},\lambda)$. That would again reduce (\ref{23}) to the sum of only the first two integrals, which is then easily seen to be bounded by $\pm2$ using (\ref{2333}). But we have again compromised the independence of the last two results to achieve this, which is not permitted. One can also easily reduce the bounds on (\ref{23}) from $\pm4$ to $\pm2$ by violating some of the rules of anti-derivatives, or of statistical averages, {\it etc.}, as an extreme measure, but that again amounts to forfeiting the game from the start. Thus, the only admissible means of reducing the bounds on (\ref{23}) from $\pm4$ to $\pm2$ is by using the standard rules of anti-derivatives that allow us to equate the sum of\break integrals appearing in (\ref{23}) to the integral (\ref{home}) of the sum of eigenvalues, which, as we saw in (\ref{not5}), is bounded by $\pm2$.

We have thus derived the additivity of expectation values (\ref{ladd}) by imposing the bounds of $\pm2$ on the Bell-CHSH sum (\ref{combi}) as our starting assumption. Therefore, given the previous derivation that led us to (\ref{chsh}) by assuming (\ref{ladd}) and the current derivation that led us to (\ref{ladd}) by assuming (\ref{chsh}), we have proved that the assumption (\ref{ladd}) of the additivity of expectation values is tautologous to assuming the bounds of $\pm2$ on the Bell-CHSH sum (\ref{combi}) of expectation values.

\subsection{Additivity of expectation values (\ref{ladd}) is an unjustified assumption, equivalent to the thesis to be proven} \label{Sec-E}

The key step that led us to the bounds of $\pm2$ on (\ref{combi}) that are more restrictive than $\pm2\sqrt{2}$ is the step (\ref{ladd}) of the linear additivity of expectation values. In what follows, I will demonstrate that this step is, in fact, an unjustified assumption, equivalent to the main thesis of the theorem to be proven, just as it is in von Neumann's now discredited theorem \cite{Oversight,Bell-1966,Grete,Mermin,BohmBub}. But this fact is obscured by the seemingly innocuous built-in linear additivity of integrals used in step (\ref{ladd}). However, as we noted around (\ref{notfl9}) and will be further demonstrated in Section~\ref{Sec-F}, the built-in linear additivity of integrals is physically meaningful only for simultaneously measurable or commuting observables \cite{Grete,Mermin}. It is, therefore, not legitimate to invoke it at step~(\ref{ladd}) {\it without proof}. Step (\ref{ladd}) would be valid also in classical physics in which the value of a sum of observable quantities would be the same as the sum of the values each quantity would take separately, because, unlike in quantum mechanics, they would all be simultaneously measurable, yielding only sharp values. Perhaps for this reason it is usually not viewed as an assumption but mistaken for a benign mathematical step. It is also sometimes claimed to be necessitated by Einstein's requirement of realism \cite{Einstein}. But I will soon explain why it is a much overlooked unjustified assumption, and demonstrate in Section~\ref{Sec-F} that, far from being required by realism, the right-hand side of step (\ref{ladd}), in fact, {\it contradicts} realism, which requires that {\it every} observable of a physical system,\break including any sums of observables, must be assigned a correct eigenvalue, quantifying one of its preexisting properties.

Moreover, realism has already been adequately accommodated by the very definition of the local functions ${\mathscr A}({\bf a},\lambda)$ and ${\mathscr B}({\bf b},\lambda)$ and their counterfactual juxtaposition on the left-hand side of (\ref{ladd}), as contextually existing properties of the system. Evidently, while a result in only one of the four expectation values corresponding to a sub-experiment that appears on the left-hand side of (\ref{ladd}) can be realized in a given run of a Bell-test experiment, the remaining three results appearing on that side are realizable at least counterfactually, thus fulfilling the requirement of realism \cite{Oversight}.~Therefore, the requirement of realism does not necessitate the left-hand side of (\ref{ladd}) to be equated with its right-hand side in the derivation of (\ref{chsh}).~Realism requires definite results ${\mathscr A}({\bf a},\lambda)\,{\mathscr B}({\bf b},\lambda)$ to exist as eigenvalues only counterfactually, {\it not all}\break {\it four at once}, as they are written on the right-hand side of (\ref{ladd}).~What is more, as we will soon see, realism implicit in the prescription (\ref{99}) requires the quantity (\ref{int}) to be a {\it correct} eigenvalue of the summed operator (\ref{op}), but it is not.

On the other hand, given the assumption $p(\lambda\,|\,{\mathbf a},{\mathbf b})=p(\lambda)$ of statistical independence and the additivity property of anti-derivatives, mathematically the equality (\ref{ladd}) follows at once because of the linearity built into the integrals, provided we adopt a double standard for additivity: we reject (\ref{ladd}) for von~Neumann's theorem as Bell did in \cite{Bell-1966}, but accept it unreservedly for Bell's theorem \cite{Oversight,Grete}. The binary properties of the functions ${\mathscr A}({\bf a},\lambda)$ and ${\mathscr B}({\bf b},\lambda)$ then immediately lead us to the bounds of $\pm2$ on (\ref{combi}). But, as we saw above, assuming the bounds of $\pm2$ on (\ref{combi}) leads, conversely, to the assumption (\ref{ladd}) of additivity of expectation values. Thus, assuming the additivity of expectation values (\ref{ladd}) is mathematically equivalent to assuming the bounds of $\pm2$ on the Bell-CHSH sum (\ref{combi}). In other words, Bell's argument presented in Section~\ref{Sec-D} {\it assumes} its conclusion (\ref{chsh}) in the guise of assumption (\ref{ladd}), by implicitly assuming that the expectation functions $(\Psi,\,\lambda\,|\,{\Omega}({c})\,|\,\Psi,\,\lambda)$ determining the eigenvalues ${\omega}({c},\lambda)$ of ${\Omega}({c})$ are {\it linear}~\cite{BohmBub}:
\begin{equation}
\widetilde{\omega}(\widetilde{c},\,\lambda)=(\,\Psi,\,\lambda\,|\left[\sum_{i=1}^4\Omega_i(c_i)\right]|\,\Psi,\,\lambda\,) =
\sum_{i=1}^4\,(\,\Psi,\,\lambda\,|\,\Omega_i(c_i)\,|\,\Psi,\,\lambda\,)=\sum_{i=1}^4\,\omega_i(c_i,\,\lambda) . \label{sumdis}
\end{equation}
But, as explained by Bohm and Bub in \cite{BohmBub} (see Appendix~\ref{C} below), this assumed linearity of $(\Psi,\,\lambda\,|\,{\Omega}({c})\,|\,\Psi,\,\lambda)$ is unreasonably restrictive for dispersion-free states $|\,\Psi,\,\lambda)$, because the observables defined in (\ref{obs}) are not simultaneously measurable. However, it allows us to reduce the following correct relation within quantum mechanics {\it as well as} hidden variable theories,
\begin{equation}
\sum_{i=1}^4\left[\int_{\mathscr L}\,(\,\Psi,\,\lambda\,|\,\Omega_i(c_i)\,|\,\Psi,\,\lambda\,)\;p(\lambda)\,d\lambda\,\right]=\int_{\mathscr L}
\,(\,\Psi,\,\lambda\,|\left[\sum_{i=1}^4\Omega_i(c_i)\right]|\,\Psi,\,\lambda\,)\;p(\lambda)\,d\lambda\,, \label{27q}
\end{equation}
to the relation
\begin{equation}
\sum_{i=1}^4\left[\int_{\mathscr L}\,(\,\Psi,\,\lambda\,|\,\Omega_i(c_i)\,|\,\Psi,\,\lambda\,)\;p(\lambda)\,d\lambda\,\right]=\int_{\mathscr L}\left[\sum_{i=1}^4\,(\,\Psi,\,\lambda\,|\,\Omega_i(c_i)\,|\,\Psi,\,\lambda\,)\right]p(\lambda)\,d\lambda\,, \label{sumdisint}
\end{equation}
which is the same as assumption (\ref{ladd}), albeit written in a more general notation.~The equality (\ref{27q}), on the other hand, is equivalent to the quantum mechanical relation (\ref{qadd}) discussed below, which can be verified using the prescription (\ref{99}).\break The same equality (\ref{27q}) is also valid for hidden variable theories, because it does not make the mistake of relying on the linearity assumption (\ref{sumdis}). This can be verified also using (\ref{99}) and the ansatz (\ref{hidres}). Thus, the innocuous-looking linear additivity of integrals in assumption (\ref{ladd}), while mathematically correct, is neither innocent nor physically reasonable.

It is not difficult to understand why appealing to the built-in linear additivity of anti-derivatives is not as innocent or physically reasonable as it may seem. In fact, for non-commuting observables that are not simultaneously measurable, justification of (\ref{ladd}) or (\ref{sumdisint}) by appealing to the built-in linear additivity of integrals leads to {\it incorrect equality between unequal physical quantities}. The reasons for this were recognized by Grete Hermann \cite{Grete} some three decades before the formulation of Bell's theorem \cite{Bell-1964}, as part of her insightful criticism of von Neumann's alleged theorem \cite{vonNeumann,Mermin}. As she explained in \cite{Grete}, we are not concerned here with classical physics in which all observable quantities are simultaneously measurable yielding only sharp values, and therefore the value of a sum of observable quantities is nothing other than the sum of the values each of those quantities would separately take. Consequently, in classical physics, the averages  of such values over individual initial states $\lambda$ of the system can also be meaningfully added linearly, just as assumed in step (\ref{ladd}) or (\ref{sumdisint}), because there is no scope for any contradiction between the averages obtained by evaluating the left-hand side and the right-hand side of these equations. Therefore, in classical physics linear additivity of expectation values remains consistent with the built-in linear additivity of anti-derivatives. However, the same cannot be assumed without proof for the dispersion-free states $|\psi,\,\lambda)$ of hidden variable theories, because, in that case, the values of the observable quantities are eigenvalues of the corresponding quantum mechanical operators dictated by the ansatz (\ref{hidres}), and, as we noted above and toward the end of Section~\ref{Sec-B}, the eigenvalue ${\widetilde{\omega}(\tilde{c},\,\lambda)}$ of the summed observable $\widetilde{\Omega}(\tilde{c})$ is not equal to the sum $\sum_{i=1}^n\omega_i(c_i,\,\lambda)$ of the eigenvalues $\omega_i(c_i,\,\lambda)$ of $\Omega_i(c_i)$, unless the observables $\Omega_i(c_i)$ constituting the sum $\widetilde{\Omega}(\tilde{c})$ are simultaneously measurable. {\it Thus, an important step in the proof of (\ref{chsh}) is missing}. A necessary step that would prove the consistency of the built-in linear additivity of anti-derivatives with the non-additivity of expectation values for the {\it non-commuting} observables. In equation (\ref{correct}) of Section$\,$\ref{Sec-F} below we will see the difference between the eigenvalue of the summed operator and the sum of individual eigenvalues explicitly. It will demonstrate how, in hidden variable theories equation (\ref{ladd}) or (\ref{sumdisint}) involving averages of eigenvalues ends up equating unequal averages of physical quantities in general. It will thereby prove that, while valid in classical physics and for simultaneously measurable observables, equation (\ref{ladd}) or (\ref{sumdisint}) is {\it not} valid for hidden variable theories in general. Insisting otherwise thus amounts to {\it assuming} the validity of this equation {\it without proof}, despite the contrary evidence just presented \cite{Grete}. That, in turn, amounts to assuming the very thesis to be proven --- namely, the bounds of $\pm2$ on the Bell-CHSH sum (\ref{combi}). Consequently, the only\break correct meaning assignable to (\ref{ladd}) or (\ref{chsh}) is that it is valid only in classical physics and/or for commuting observables.

Sometimes assumption (\ref{ladd}) is justified on statistical grounds. It is argued that the four sub-experiments appearing on the left-hand side of (\ref{ladd}) with different experimental settings $\{{\bf a},\,{\bf b}\}$, $\{{\bf a},\,{\bf b'}\}$, {\it etc.} can be performed independently of each other, on possibly different occasions, and then the resulting averages are added together at a later time for statistical analysis. If the number of experimental runs for each pair of settings is sufficiently large, then, theoretically, the sum of the four averages appearing on the left-hand side of (\ref{ladd}) are found not to exceed the bounds of $\pm2$, thus justifying the equality (\ref{ladd}). This can be easily verified in numerical simulations (see Ref.~[27] cited in \cite{IEEE-4}). However,\break this heuristic argument is not an analytical proof of the bounds. What it implicitly neglects to take into account by explicitly assuming that the four sub-experiments can be performed independently, is that the sub-experiments involve mutually exclusive pairs of settings such as $\{{\bf a},\,{\bf b}\}$ and $\{{\bf a},\,{\bf b'}\}$ in physical space, and thus involve non-commuting observables that cannot be measured simultaneously \cite{Oversight}. Unless the statistical analysis takes this physical fact into account, it cannot be claimed to have any relevance for the Bell-test experiments \cite{RSOS-Reply}. For ignoring this physical fact amounts to incorrectly assuming that the spin observables ${\boldsymbol\sigma}_1\cdot{\bf a}\,\otimes\,{\boldsymbol\sigma}_2\cdot{\bf b}$, {\it etc.} are mutually commuting, and thus simultaneously measurable, for which assumption (\ref{ladd}) is indeed valid, as demonstrated below in Section~\ref{Sec-F} (see the discussion around (\ref{incorrect})). On the other hand, when the non-commutativity of the observables involved in the sub-experiments is taken into account in numerical simulations, the bounds on (\ref{combi}) turn out to be $\pm2\sqrt{2}$, as shown in \cite{RSOS,IEEE-2} and Ref.~[27] cited in \cite{IEEE-4}. In other words, such a statistical argument is simply assumption (\ref{ladd}) in disguise.

Another important point to recognize here is that the above derivation of the stringent bounds of $\pm2$ on (\ref{combi}) for a locally causal dispersion-free counterpart $|\,\Psi,\,\lambda)$ of the quantum mechanical singlet state (\ref{single}) must comply with the heuristics of the contextual hidden variable theories we discussed in Section~\ref{Sec-B}. If it does not, then the bounds of $\pm2$ cannot be claimed to have any relevance for the viability of local hidden variable theories \cite{Shimony}. Therefore, as discussed in Section~\ref{Sec-B}, in a contextual hidden variable theory all of the observables $\Omega_i(c_i)$ of any physical system, {\it including} their sum $\widetilde{\Omega}(\tilde{c})=\sum_{i=1}^n\Omega_i(c_i)$, which also represents a physical quantity in the Hilbert space formulation of quantum mechanics \cite{vonNeumann} whether or not it is observed, must be assigned unique eigenvalues $\omega_i(c_i,\,\lambda)$ and $\widetilde{\omega}(\tilde{c},\,\lambda)$, respectively, in the dispersion-free states $|\,\psi,\,\lambda)$ of the system, regardless of whether these observables are simultaneously measurable. In particular, while the summed observable (\ref{op}) discussed below is never observed in the Bell-test experiments, realism nevertheless requires it to be assigned a unique eigenvalue in accordance with the ansatz (\ref{hidres}) and the prescription~(\ref{99}).

\subsection{Additivity of expectation values is respected by quantum states} \label{Sec-E2}

Now, within quantum mechanics, expectation values do add in analogy with the equality (\ref{ladd}) assumed by Bell for local hidden variable theories \cite{vonNeumann,Bell-1966}. In quantum mechanics, the statistical predictions of which any hidden variable theory is obliged to reproduce, the joint results ${\mathscr A}({\bf a},\,\lambda)\,{\mathscr B}({\bf b},\,\lambda)$ observed by Alice and Bob would be eigenvalues of the operators ${\boldsymbol\sigma}_1\cdot{\bf a}\,\otimes\,{\boldsymbol\sigma}_2\cdot{\bf b}$, and the linearity in the rules of Hilbert space quantum mechanics ensures that these operators satisfy the additivity of expectation values.~Thus, for any quantum state $|\psi\rangle$, the following equality holds:
\begin{align}
\langle\psi|\,{\boldsymbol\sigma}_1\cdot{\bf a}\,&\otimes\,{\boldsymbol\sigma}_2\cdot{\bf b}\,|\psi\rangle+\langle\psi|\,{\boldsymbol\sigma}_1\cdot{\bf a}\,\otimes\,{\boldsymbol\sigma}_2\cdot{\bf b'}\,|\psi\rangle+\langle\psi|\,{\boldsymbol\sigma}_1\cdot{\bf a'}\,\otimes\,{\boldsymbol\sigma}_2\cdot{\bf b}\,|\psi\rangle-\langle\psi|\,{\boldsymbol\sigma}_1\cdot{\bf a'}\,\otimes\,{\boldsymbol\sigma}_2\cdot{\bf b'}\,|\psi\rangle \notag \\[3pt]
&=\,\langle\psi|\,{\boldsymbol\sigma}_1\cdot{\bf a}\,\otimes\,{\boldsymbol\sigma}_2\cdot{\bf b}+{\boldsymbol\sigma}_1\cdot{\bf a}\,\otimes\,{\boldsymbol\sigma}_2\cdot{\bf b'}+{\boldsymbol\sigma}_1\cdot{\bf a'}\,\otimes\,{\boldsymbol\sigma}_2\cdot{\bf b}-{\boldsymbol\sigma}_1\cdot{\bf a'}\,\otimes\,{\boldsymbol\sigma}_2\cdot{\bf b'}\,|\psi\rangle. \label{qadd}
\end{align}
The linearity of this equation has been achieved by promoting observable quantities to Hermitian operators \cite{vonNeumann,Grete}. Any local hidden variable theory is then obliged to reproduce the predictions of both sides of this equation. Comparing equations (\ref{ladd}) and (\ref{qadd}), the equality between the two sides of (\ref{ladd}) seems reasonable, even physically. Furthermore, since the condition (\ref{first}) for any hidden variable theory obliges us to set the four terms on the left-hand side of (\ref{qadd}) as
\begin{subequations}
\begin{align}
\langle\Psi|\,{\boldsymbol\sigma}_1\cdot{\bf a}\,\otimes\,{\boldsymbol\sigma}_2\cdot{\bf b}\,|\Psi\rangle&=\!\int_{\mathscr L}{\mathscr A}({\mathbf a},\,\lambda)\,{\mathscr B}({\mathbf b},\,\lambda)\;p(\lambda)\,d\lambda\,, \\
\langle\Psi|\,{\boldsymbol\sigma}_1\cdot{\bf a}\,\otimes\,{\boldsymbol\sigma}_2\cdot{\bf b'}\,|\Psi\rangle&=\!\int_{\mathscr L}{\mathscr A}({\mathbf a},\,\lambda)\,{\mathscr B}({\mathbf b'},\,\lambda)\;p(\lambda)\,d\lambda\,, \\
\langle\Psi|\,{\boldsymbol\sigma}_1\cdot{\bf a'}\,\otimes\,{\boldsymbol\sigma}_2\cdot{\bf b}\,|\Psi\rangle&=\!\int_{\mathscr L}{\mathscr A}({\mathbf a'},\,\lambda)\,{\mathscr B}({\mathbf b},\,\lambda)\;p(\lambda)\,d\lambda\,, \\
\text{and}\;\;\langle\Psi|\,{\boldsymbol\sigma}_1\cdot{\bf a'}\,\otimes\,{\boldsymbol\sigma}_2\cdot{\bf b'}\,|\Psi\rangle&=\!\int_{\mathscr L}{\mathscr A}({\mathbf a'},\,\lambda)\,{\mathscr B}({\mathbf b'},\,\lambda)\;p(\lambda)\,d\lambda\,,
\end{align}
\end{subequations}
it may seem reasonable that, given the quantum mechanical equality (\ref{qadd}), any hidden variable theory should satisfy
\begin{align}
\langle\Psi|\,\widetilde{\Omega}(\tilde{c})\,|\Psi\rangle&=\langle\Psi|\,{\boldsymbol\sigma}_1\cdot{\bf a}\,\otimes\,{\boldsymbol\sigma}_2\cdot{\bf b}+{\boldsymbol\sigma}_1\cdot{\bf a}\,\otimes\,{\boldsymbol\sigma}_2\cdot{\bf b'}+{\boldsymbol\sigma}_1\cdot{\bf a'}\,\otimes\,{\boldsymbol\sigma}_2\cdot{\bf b}-{\boldsymbol\sigma}_1\cdot{\bf a'}\,\otimes\,{\boldsymbol\sigma}_2\cdot{\bf b'}\,|\Psi\rangle \notag \\
&=\!\int_{\mathscr L}\!\!\big\{\,{\mathscr A}({\bf a},\lambda)\,{\mathscr B}({\bf b},\lambda)+{\mathscr A}({\bf a},\lambda)\,{\mathscr B}({\bf b'},\lambda)+{\mathscr A}({\bf a'},\lambda)\,{\mathscr B}({\bf b},\lambda)-{\mathscr A}({\bf a'},\lambda)\,{\mathscr B}({\bf b'},\lambda)\big\}\;p(\lambda)\,d\lambda\,,\label{wrong}
\end{align}
adhering to the prescription (\ref{99}), which would then justify equality (\ref{ladd}). Since hidden variable theories are required to satisfy the prescription (\ref{99}), should not they also reproduce equation (\ref{wrong})?~The answer to this is not straightforward.

\subsection{Additivity of expectation values does not hold for dispersion-free states} \label{Sec-F}

The problem with equation (\ref{wrong}) is that, while the joint results ${\mathscr A}(\mathbf{a},\lambda){\mathscr B}(\mathbf{b},\lambda)$, {\it etc.} appearing on the left-hand side of equation (\ref{ladd}) are possible eigenvalues of the products of spin operators ${\boldsymbol\sigma}_1\cdot{\bf a}\,\otimes\,{\boldsymbol\sigma}_2\cdot{\bf b}$, {\it etc.}, their summation
\begin{equation}
{\mathscr A}({\bf a},\,\lambda)\,{\mathscr B}({\bf b},\,\lambda)+{\mathscr A}({\bf a},\,\lambda)\,{\mathscr B}({\bf b'},\,\lambda)+{\mathscr A}({\bf a'},\,\lambda)\,{\mathscr B}({\bf b},\,\lambda)-{\mathscr A}({\bf a'},\,\lambda)\,{\mathscr B}({\bf b'},\,\lambda) \label{bell}
\end{equation}
appearing as the integrand on the right-hand side of equation (\ref{wrong}) or (\ref{ladd}) is {\it not} an eigenvalue of the summed operator
\begin{equation}
\widetilde{\Omega}(\tilde{c})={\boldsymbol\sigma}_1\cdot{\bf a}\,\otimes\,{\boldsymbol\sigma}_2\cdot{\bf b}+{\boldsymbol\sigma}_1\cdot{\bf a}\,\otimes\,{\boldsymbol\sigma}_2\cdot{\bf b'}+{\boldsymbol\sigma}_1\cdot{\bf a'}\,\otimes\,{\boldsymbol\sigma}_2\cdot{\bf b}-{\boldsymbol\sigma}_1\cdot{\bf a'}\,\otimes\,{\boldsymbol\sigma}_2\cdot{\bf b'}, \label{op}
\end{equation}
because the spin operators ${\boldsymbol\sigma}_1\cdot{\bf a}$ and ${\boldsymbol\sigma}_1\cdot{\bf a'}$, {\it etc.}, and therefore ${\boldsymbol\sigma}_1\cdot{\bf a}\,\otimes\,{\boldsymbol\sigma}_2\cdot{\bf b}$, {\it etc.}, do not commute with each other:
\begin{align}
\left[\,{\boldsymbol\sigma}_1\cdot{\bf a}\,\otimes\,{\boldsymbol\sigma}_2\cdot{\bf b},\;{\boldsymbol\sigma}_1\cdot{\bf a}\,\otimes\,{\boldsymbol\sigma}_2\cdot{\bf b'}\,\right]&=
2\,{\boldsymbol\sigma}\cdot\left\{\left({\bf a}\times{\bf b'}\right)\times\left({\bf a}\times{\bf b}\right)\right\} \notag \\
&\not=0 \;\,\text{if}\;\,
{\bf b'}\not={\bf b}\not={\bf a}. \label{nop}
\end{align}
Consequently, equation (\ref{wrong}) would hold within any hidden variable theory {\it only if} the operators ${\boldsymbol\sigma}_1\cdot{\bf a}\,\otimes\,{\boldsymbol\sigma}_2\cdot{\bf b}$, {\it etc.}, were commuting operators. As we discussed, this is well known from the famous criticisms of von Neumann's theorem against hidden variable theories \cite{Oversight,Bell-1966,Grete,Mermin}. While the equality (\ref{ladd}) of the sum of expectation values with the expectation value of the sum is respected in quantum mechanics, it does not hold for hidden variable theories \cite{Bell-1966}. Nor does local realism necessitate the linear additivity (\ref{sumdis}) of eigenvalues for individual dispersion-free states $|\,\Psi,\,\lambda)$.

This problem, however, suggests its own resolution.~We can work out the correct eigenvalue $\widetilde{\omega}(\tilde{c},\,\lambda)$ of the summed operator (\ref{op}), at least formally, as I have worked out in Appendix~\ref{A} below.~The correct version of equation (\ref{wrong})~is~then
\begin{equation}
\langle\Psi|\,{\boldsymbol\sigma}_1\cdot{\bf a}\,\otimes\,{\boldsymbol\sigma}_2\cdot{\bf b}+{\boldsymbol\sigma}_1\cdot{\bf a}\,\otimes\,{\boldsymbol\sigma}_2\cdot{\bf b'}+{\boldsymbol\sigma}_1\cdot{\bf a'}\,\otimes\,{\boldsymbol\sigma}_2\cdot{\bf b}-{\boldsymbol\sigma}_1\cdot{\bf a'}\,\otimes\,{\boldsymbol\sigma}_2\cdot{\bf b'}\,|\Psi\rangle=\!\int_{\mathscr L} {\widetilde{\omega}}({\bf a},{\bf a'},{\bf b},{\bf b'},\lambda)\;p(\lambda)\,d\lambda\,, \label{corsum}
\end{equation}
where
\begin{equation}
{\widetilde{\omega}}\!=\pm\sqrt{\big\{{\mathscr A}({\bf a},\,\lambda)\,{\mathscr B}({\bf b},\,\lambda)+{\mathscr A}({\bf a},\,\lambda)\,{\mathscr B}({\bf b'},\,\lambda)+{\mathscr A}({\bf a'},\,\lambda)\,{\mathscr B}({\bf b},\,\lambda)-{\mathscr A}({\bf a'},\,\lambda)\,{\mathscr B}({\bf b'},\,\lambda) \big\}^2 + (\Psi,\,\lambda\,|\,{\widetilde{\Theta}}\,|\,\Psi,\,\lambda)\,} \label{correct}
\end{equation}
is the correct eigenvalue of the summed operator (\ref{op}), with its non-commuting part separated out as the operator
\begin{equation}
{\widetilde{\Theta}}({\bf a},{\bf a'},{\bf b},{\bf b'})=2\,{\boldsymbol\sigma}\cdot{\bf n}({\bf a},{\bf a'},{\bf b},{\bf b'})\,. \label{theta}
\end{equation}
Here $(\Psi,\,\lambda\,|\,{\widetilde{\Theta}}\,|\,\Psi,\,\lambda)\not=0$ in general, because the vector ${\bf n}({\bf a},{\bf a'},{\bf b},{\bf b'})$ does not vanish in general. It works out to be 
\begin{align}
{\bf n}({\bf a},{\bf a'},{\bf b},{\bf b'})=\big\{\left({\bf a}\times{\bf b'}\right)\times\left({\bf a}\times{\bf b}\right)&+\left({\bf a'}\times{\bf b}\right)\times\left({\bf a}\times{\bf b}\right)+\left({\bf a'}\times{\bf b}\right)\times\left({\bf a}\times{\bf b'}\right) \notag \\
&-\left({\bf a'}\times{\bf b'}\right)\times\left({\bf a}\times{\bf b}\right)-\left({\bf a'}\times{\bf b'}\right)\times\left({\bf a'}\times{\bf b}\right)-\left({\bf a'}\times{\bf b'}\right)\times\left({\bf a}\times{\bf b'}\right)\big\}. \label{vec}
\end{align}
The details of how this separation is accomplished using (\ref{nop}) can be found in Appendix~\ref{A} below.~From (\ref{correct}), it is now easy to appreciate that the additivity of expectation values (\ref{ladd}) assumed by Bell can hold only if the expectation value $(\Psi,\lambda\,|\,{\widetilde{\Theta}}\,|\,\Psi,\lambda)=\pm2\,||{\bf n}||$ of the non-commuting part within the eigenvalue ${\widetilde{\omega}}({\bf a},{\bf a'},{\bf b},{\bf b'},\lambda)$ of the summed operator (\ref{op}) is zero. But that is possible only if the operators ${\boldsymbol\sigma}_1\cdot{\bf a}\,\otimes\,{\boldsymbol\sigma}_2\cdot{\bf b}$, {\it etc.} constituting the sum (\ref{op}) commute with each\break other.~In general, if the operators ${\boldsymbol\sigma}_1\cdot{\bf a}\,\otimes\,{\boldsymbol\sigma}_2\cdot{\bf b}$, {\it etc.} in (\ref{op}) do not commute with each other, then we would have
\begin{equation}
{\widetilde{\omega}}({\bf a},{\bf a'},{\bf b},{\bf b'},\lambda)\not={\mathscr A}({\bf a},\,\lambda)\,{\mathscr B}({\bf b},\,\lambda)+{\mathscr A}({\bf a},\,\lambda)\,{\mathscr B}({\bf b'},\,\lambda)+{\mathscr A}({\bf a'},\,\lambda)\,{\mathscr B}({\bf b},\,\lambda)-{\mathscr A}({\bf a'},\,\lambda)\,{\mathscr B}({\bf b'},\,\lambda). \label{incorrect}
\end{equation}
But the operators ${\boldsymbol\sigma}_1\cdot{\bf a}\,\otimes\,{\boldsymbol\sigma}_2\cdot{\bf b}$, {\it etc.} indeed do not commute with each other, because the pairs of directions $\{{\bf a},\,{\bf a'}\}$, {\it etc.} in (\ref{op}) are mutually exclusive directions in ${\mathrm{I\!R}^3}$.~Therefore, the additivity of expectation values assumed at step (\ref{ladd}) in the derivation of (\ref{chsh}) is unjustifiable. Far from being necessitated by realism, it actually contradicts realism.

Since three of the four results appearing in the expression (\ref{bell}) can be realized only counterfactually, their summation in (\ref{bell}) cannot be realized {\it even} counterfactually \cite{Oversight}. Thus, in addition to not being a correct eigenvalue of the summed operator (\ref{op}) as required by the prescription (\ref{99}) for hidden variable theories, the quantity appearing in (\ref{bell}) is, in fact, an entirely fictitious quantity, with no counterpart in any possible world, apart from in the trivial case when all observables are commutative. By contrast, the correct eigenvalue (\ref{correct}) of the summed operator (\ref{op}) can be realized at least counterfactually because it is a genuine eigenvalue of that operator, thereby satisfying the requirement of realism correctly, in accordance with the prescription (\ref{99}) for hidden variable theories. Using (\ref{correct}), all five of the observables appearing on both sides of the quantum mechanical equation (\ref{qadd}) can be assigned unique and correct eigenvalues \cite{Oversight}.

Once this oversight is ameliorated, it is not difficult to show that the conclusion of Bell's theorem no longer follows. For then, using the correct eigenvalue (\ref{correct}) of (\ref{op}) instead of (\ref{bell}) on the right-hand side of (\ref{ladd}), we have the equation
\begin{equation}
{\cal E}({\bf a},\,{\bf b})+{\cal E}({\bf a},\,{\bf b'})+{\cal E}({\bf a'},\,{\bf b})-{\cal E}({\bf a'},\,{\bf b'})=\!\int_{\mathscr L} {\widetilde{\omega}}({\bf a},{\bf a'},{\bf b},{\bf b'},\lambda)\;p(\lambda)\,d\lambda\,\label{corbon}
\end{equation}
instead of (\ref{ladd}), which implements local realism correctly on both of its sides, as required by the prescription (\ref{99}) we discussed in Section~\ref{Sec-B}.~This equation (\ref{corbon}) is thus the correct dispersion-free counterpart of the equivalence (\ref{qadd}) for the quantum mechanical expectation values \cite{Oversight}. It can reduce to Bell's assumption (\ref{ladd}) only when the expectation value $(\Psi,\lambda\,|\,{\widetilde{\Theta}}\,|\,\Psi,\lambda)$ of the non-commuting part within the eigenvalue ${\widetilde{\omega}}({\bf a},{\bf a'},{\bf b},{\bf b'},\lambda)$ of the summed operator (\ref{op}) happens to be vanishing. It thus expresses the correct relationship (\ref{27q}) among the expectation values for the singlet state (\ref{single}) in the local hidden variable framework considered by Bell \cite{Bell-1964}.~Recall again from the end of Section~\ref{Sec-B} that the quantum mechanical relation (\ref{qadd}) is an unusual property of the quantum states $|\psi\rangle$.~As Bell stressed in \cite{Bell-1966}, ``[t]here is no reason to demand it individually of the hypothetical dispersion free states, whose function it is to reproduce the {\it measurable} peculiarities of quantum mechanics {\it when averaged over}.''~Moreover, in Section~V of \cite{Oversight} I have demonstrated that the bounds on the right-hand side of (\ref{corbon}) are $\pm2\sqrt{2}$ instead of $\pm2$. An alternative derivation of these bounds follows from the magnitude $||{\bf n}||$ of the vector defined in (\ref{vec}), which, as proved in Appendix~\ref{B} below, is bounded by $2$, and therefore\break the eigenvalue $\pm2\,||{\bf n}||$ of the operator (\ref{theta}) obtained as its expectation value $(\Psi,\lambda\,|\,{\widetilde{\Theta}}\,|\,\Psi,\lambda)$ is bounded by $\pm4$, giving
\begin{equation}
-4\,\leqslant(\Psi,\lambda\,|\,{\widetilde{\Theta}}({\bf a},{\bf a'},{\bf b},{\bf b'})\,|\,\Psi,\lambda)\leqslant+4\,.
\end{equation}
Substituting these into (\ref{correct}), together with the bounds of $\pm2$ we worked out before on the commuting part (\ref{bell}), gives 
\begin{equation}
-2\sqrt{2}\,\leqslant\,{\widetilde{\omega}}({\bf a},{\bf a'},{\bf b},{\bf b'},\lambda)\leqslant+2\sqrt{2}\,,
\end{equation}
which is constrained to be real despite the square root in the expression (\ref{correct}) because the operator (\ref{op}) is Hermitian. Consequently, we obtain the following Tsirel'son's bounds in the dispersion-free state, on the right-hand side of (\ref{corbon}): 
\begin{equation}
-2\sqrt{2}\,\leqslant\int_{\mathscr L}{\widetilde{\omega}}({\bf a},{\bf a'},{\bf b},{\bf b'},\lambda)\;p(\lambda)\,d\lambda\,\leqslant+2\sqrt{2}\,.
\end{equation}
Given the correct relation (\ref{corbon}) between expectation values instead of the flawed assumption (\ref{ladd}), we thus arrive at
\begin{equation}
-2\sqrt{2}\,\leqslant\,{\cal E}({\bf a},\,{\bf b})+{\cal E}({\bf a},\,{\bf b'})+{\cal E}({\bf a'},\,{\bf b})-{\cal E}({\bf a'},\,{\bf b'})\leqslant+2\sqrt{2}\,. \label{44final}
\end{equation}
Since the bounds of $\pm2\sqrt{2}$ we have derived on the Bell-CHSH sum of expectation values are the same as those predicted by quantum mechanics and observed in the Bell-test experiments, the conclusion of Bell's theorem is mitigated. What is ruled out by these experiments is not local realism but the assumption of the additivity of expectation values, which does not hold for non-commuting observables in dispersion-free states of any hidden variable theories to begin with.

It is also instructive to note that the intermediate bounds $\pm2\sqrt{2}$ on the Bell-CHSH sum (\ref{combi}), instead of the extreme bounds $\pm2$ or $\pm4$, follow in the above derivation of (\ref{44final}) as a consequence of the geometry of physical space \cite{IEEE-1,Christian}. Thus, what is brought out in it is the oversight of the non-commutative or Clifford-algebraic attributes of the physical space in Bell's derivation of the bounds $\pm2$ in (\ref{chsh}). Indeed, it is evident from Appendix~\ref{B} below that the geometry of physical space imposes the bounds $0\leqslant||{\bf n}||\leqslant2$ on the magnitude of the vector (\ref{vec}), which, in turn, lead us to the bounds $\pm2\sqrt{2}$ in (\ref{44final}). This is in sharp contrast with the traditional view of these bounds as due to non-local influences,\break stemming from a failure of the locality condition (\ref{fact}). But in the derivation of (\ref{44final}) above, the condition (\ref{fact}) is strictly respected. Therefore, the strength of the bounds $\pm2\sqrt{2}$ in (\ref{44final}) is a consequence --- {\it not} of non-locality or non-reality, but of the geometry of physical space \cite{IEEE-1,Christian}. Non-locality or non-reality is necessitated only if one erroneously insists on linear additivity (\ref{ladd}) of eigenvalues of non-commuting observables for each {\it individual} dispersion-free state $|\,\Psi,\,\lambda)$.

\subsection{Oversight and circular reasoning in the GHZ variant of Bell's theorem}\label{GHZ-flaw}

As is well known, Bell's theorem \cite{Bell-1964} has inspired several variant arguments against hidden variable theories. Some of these arguments, such as that by Clauser and Horne \cite{CH}, involve inequalities similar to the Bell-CHSH inequalities, resulting from additions of expectation values. They are therefore as flawed as Bell's original argument, and for the same reasons we discussed previously. However, the variant argument by Greenberger, Horne, and Zeilinger (GHZ) \cite{GHZ} is significantly different. It neither involves inequalities nor relies on additions of expectation values. It purports to demonstrate a purely algebraic incompatibility between the quantum mechanical predictions and the premisses of the argument by Einstein, Podolsky, and Rosen (EPR) \cite{EPR} for their program of completing quantum mechanics. Namely, their premisses of locality, reality, completeness, and perfect correlation. However, in this section I demonstrate that, despite appearances, the GHZ argument is just as flawed as Bell's original argument, and for nearly the same reasons.

\subsubsection{Expectation functions $\left(\,\psi,\,\lambda\,|\,\Omega\,|\,\psi,\,\lambda\,\right) $ for non-commuting observables cannot be multiplicative}

To this end, let us begin with the observation that the oversight that invalidates both von~Neumann's theorem \cite{vonNeumann} and Bell's theorem involves incorrect use of the {\it Functional Composition Principle}, formalized by Kochen and Specker to mirror the algebraic structure of the self-adjoint operators $\Omega$ into the algebraic structure of their eigenvalues $\omega$ \cite{Kochen}:
\begin{center}
If $f:\mathrm{I\!R}\rightarrow\mathrm{I\!R}$ is a real-valued function such that $\Omega_1=f(\Omega_2)$, then $\left\langle\,\psi\,|\,\Omega_1\,|\,\psi\,\right\rangle=f(\left\langle\,\psi\,|\,\Omega_2\,|\,\psi\,\right\rangle)$,
\end{center}
for any two self-adjoint operators $\Omega_1$ and $\Omega_2$ measured in a state $\left|\,\psi\,\right\rangle$, which may or may not be a dispersion-free state.\break If it is a dispersion-free state $\left|\,\psi,\,\lambda\,\right)$, then ansatz (\ref{hidres}) simplifies the {\it Functional Composition Principle} to the following:
\begin{center}
If $f:\mathrm{I\!R}\rightarrow\mathrm{I\!R}$ is a real-valued function such that $\Omega_1=f(\Omega_2)$, then $\omega_1(\lambda)=f(\omega_2(\lambda))$,
\end{center}
where $\omega_1(\lambda)$ and $\omega_2(\lambda)$ are possible eigenvalues of $\Omega_1$ and $\Omega_2$, respectively. The rationale behind this principle is that a hidden variable theory should preserve, not only the statistical attributes of quantum mechanics captured by (\ref{77}) but also the algebraic structure of the quantum mechanical observables. Since this principle is satisfied for the statistical mechanics underlying thermodynamics, it is expected to hold in any ensemble interpretation of quantum mechanics. Moreover, the proof of Kochen-Specker theorem \cite{Kochen} that rules out non-contextual hidden variable theories depends on this principle and respects the following two important consequences that follow from it for {\it commuting} operators:
\begin{equation}
\text{\underbar{Sum Rule}}:\;\;\;\left(\,\psi,\,\lambda\,|\,\Omega_1+\,\Omega_2\,|\,\psi,\,\lambda\,\right) =\left(\,\psi,\,\lambda\,|\,\Omega_1\,|\,\psi,\,\lambda\,\right) + \left(\,\psi,\,\lambda\,|\,\Omega_2\,|\,\psi,\,\lambda\,\right),\;\text{provided}\;\left[ \Omega_1,\,\Omega_2\right]=0,\label{sumrule}
\end{equation}
and
\begin{equation}
\text{\underbar{Product Rule}}:\;\;\;\;\left(\,\psi,\,\lambda\,|\,\Omega_1\,\Omega_2\,|\,\psi,\,\lambda\,\right) =\left(\,\psi,\,\lambda\,|\,\Omega_1\,|\,\psi,\,\lambda\,\right)\left(\,\psi,\,\lambda\,|\,\Omega_2\,|\,\psi,\,\lambda\,\right),\;\text{provided}\;\left[ \Omega_1,\,\Omega_2\right]=0. \label{productrule}
\end{equation}
It is also worth recalling that a function $f:\mathrm{I\!R}\rightarrow\mathrm{I\!R}$ is said to be multiplicative if $f(x\,y)=f(x)\,f(y)$ for any $x,y\in\mathrm{I\!R}$.  

In previous sections, we have extensively discussed the linearity implicit in the Sum Rule (\ref{sumrule}). In a dispersion-free state $\left|\,\psi,\,\lambda\,\right)$, using the ansatz (\ref{hidres}), this can be expressed in terms of eigenvalues, as proved in the Appendix~\ref{C} below:
\begin{equation}
\widetilde{\omega}(\lambda) = \,\omega_1(\lambda) + \,\omega_2(\lambda),\;\text{provided}\;\left[ \Omega_1,\,\Omega_2\right]=0, \label{sumr2}
\end{equation}
where $\widetilde{\omega}(\lambda)$ is one of the eigenvalues of the summed operator $\Omega_1+\Omega_2\,$, and $\omega_1(\lambda)$ and $\omega_2(\lambda)$ are possible eigenvalues of the operators $\Omega_1$ and $\Omega_2$, respectively. As we noted above in passing, von~Neumann mistakenly assumed the linearity of this rule to hold even for non-commuting operators \cite{vonNeumann}, and, for this reason, his theorem was criticized by Bell \cite{Bell-1966} and many other authors \cite{Oversight}. However, von~Neumann did not make the mistake of assuming the Product Rule (\ref{productrule}) for non-commuting operators in his theorem, as acknowledged by Kochen and Specker \cite{Kochen}. Unfortunately, Greenberger, Horne, and Zeilinger make {\it that} mistake in their variant of Bell's theorem \cite{GHZ}. They implicitly assume {\it multiplicative} expectation function (\ref{productrule}) for {\it non-commuting} observables in their analysis to incorrectly claim that they have found a\break contradiction in the premisses of the argument by Einstein, Podolsky, and Rosen \cite{EPR}, as I now proceed to demonstrate.

Although the Product Rule (\ref{productrule}) goes back to the pioneering investigations by von~Neumann \cite{vonNeumann} and is universally accepted, since it plays a crucial role in the proof of GHZ's argument, let me demonstrate that {\it it does not} hold for non-commuting operators. To this end, using ansatz (\ref{hidres}) it is easy to see that for dispersion-free states (\ref{productrule}) simplifies~to
\begin{equation}
\widetilde{\omega}(\lambda) = \,\omega_1(\lambda)\;\omega_2(\lambda),\;\text{provided}\;\left[ \Omega_1,\,\Omega_2\right]=0, \label{prod2}
\end{equation}
where $\widetilde{\omega}(\lambda)$ is one of the eigenvalues of the product operator $\Omega_1\,\Omega_2\,$, and $\omega_1(\lambda)$ and $\omega_2(\lambda)$ are, respectively, possible eigenvalues of the individual operators $\Omega_1$ and $\Omega_2$. Now, we borrow from Bell \cite{Bell-1966} the example of a spin-$\frac{1}{2}$ particle, which he used to explain the invalidity of the Sum Rule (\ref{sumr2}) for the eigenvalues of non-commuting operators (recall the discussion above from the last paragraph of Section~\ref{Sec-B}). We can similarly demonstrate the invalidity of the Product Rule (\ref{prod2}) for the eigenvalues of non-commuting operators \cite{Kochen}. Thus, since the operators $\sigma_x$ and $\sigma_y$ do not commute, $\left[\sigma_x,\,\sigma_y\right]\not=0$, according to (\ref{prod2}) the eigenvalues of the product $\sigma_x\sigma_y$ cannot be equal to the product of the eigenvalues of $\sigma_x$ and $\sigma_y$. And, indeed, we see that the eigenvalues of $\sigma_x$ and $\sigma_y$ are both $\pm1$, whereas the eigenvalues of $\sigma_x\sigma_y=i\sigma_z$ are $\pm\,i$, and thus not even real. Therefore, as expected, the Product Rule does not hold, since $(\pm1)(\pm1)=\pm1\not=\pm\,i$. Note that what is asserted by (\ref{prod2}) is not that $\omega_1(\lambda)$, $\omega_2(\lambda)$, and $\widetilde{\omega}(\lambda)$ cannot exist simultaneously as required by realism,\break but rather that the relationship among the three cannot be multiplicative if the operators $\Omega_1$ and $\Omega_2$ do not commute.

In summary, as we discussed in the previous sections, just as the eigenvalue of a sum of observables is {\it not} equal to the sum of eigenvalues of observables constituting the sum when the latter do not commute, the eigenvalue of a product of observables is {\it not} equal to the product of eigenvalues of observables constituting the product when the latter do not commute. The conceptual and physical reasons for this are the same as those recognized by Grete Hermann \cite{Grete} (and later by Bell \cite{Bell-1966} and others \cite{Oversight}) in the context of a similar mistake in von~Neumann's pioneering theorem against hidden variable theories, as we discussed above in Sections~\ref{Sec-B} and \ref{Sec-E}. Namely, in hidden variable theories, we are not concerned\break with classical physics --- we are concerned with relationships among the eigenvalues of quantum mechanical operators.

\subsubsection{Four-particle Greenberger--Horne--Zeilinger state and the associated observables}

Now, together with Shimony, in \cite{GHZ} Greenberger, Horne, and Zeilinger considered the following four-particle state:
\begin{equation}
|\Psi_{\bf z}\rangle\,=\,\frac{1}{\sqrt{2}\,}\,\Bigl\{|{\bf z},\,+\rangle_1\otimes
|{\bf z},\,+\rangle_2\otimes|{\bf z},\,-\rangle_3\otimes
|{\bf z},\,-\rangle_4\,
-\,|{\bf z},\,-\rangle_1\otimes|{\bf z},\,-\rangle_2\otimes
|{\bf z},\,+\rangle_3\otimes|{\bf z},\,+\rangle_4\Bigr\}.\label{ghz-single}
\end{equation}
Unlike the singlet state (\ref{single}), this entangled state, composed of four fermionic particles, is not rotationally invariant. There is a privileged direction, and it is taken to be the ${\bf z}$-direction of their experimental setup. The ${\bf z}$-direction thus represents the axis of anisotropy in the system. The GHZ observables are then a direct product of four Pauli matrices:
\begin{equation}
\Omega(c)=\,{\boldsymbol\sigma}\cdot{\bf n}_1\,\otimes\,
{\boldsymbol\sigma}\cdot{\bf n}_2\,\otimes\,{\boldsymbol\sigma}\cdot{\bf n}_3\,\otimes\,{\boldsymbol\sigma}\cdot{\bf n}_4\,,\label{fullop}
\end{equation}
where the vectors ${\bf n}_1\not={\bf n}_2\not={\bf n}_3\not={\bf n}_4$ in general. Consequently, with alternative directions such as ${\bf n}'_1\not={\bf n}'_2\not={\bf n}'_3\not={\bf n}'_4$ the corresponding operator will not commute with the one expressed in (\ref{fullop}), just as in the two-particle case (\ref{nop}) we discussed in Section~\ref{Sec-F}. The quantum mechanical expectation value of the product of the four outcomes of the spin measurements, {\it i.e.}, the product of finding a spin value of particle 1 along ${\bf n}_1$, of particle 2 along ${\bf n}_2$, {\it etc.}, is given~by
\begin{equation}
{\cal E}^{\Psi_{\bf z}}_{{\!}_{QM}\!}({\bf n}_1,\,{\bf n}_2,\,{\bf n}_3,\,{\bf n}_4)\,=\,
\langle\Psi_{\bf z}|\,{\boldsymbol\sigma}\cdot{\bf n}_1\,\otimes\,
{\boldsymbol\sigma}\cdot{\bf n}_2\,\otimes\,{\boldsymbol\sigma}\cdot{\bf n}_3\,\otimes\,
{\boldsymbol\sigma}\cdot{\bf n}_4\,|\Psi_{\bf z}\rangle.\label{realobserve}
\end{equation}
This expectation function has been worked out in Appendix~F of \cite{GHZ}. In the spherical coordinates, with angles such as ${\theta_1}$ and ${\phi_1}$ representing, respectively, the polar and azimuthal angles of the directions ${\bf n}_1$, {\it etc.}, it works out to be
\begin{equation}
{\cal E}^{\Psi_{\bf z}}_{{\!}_{QM}\!}({\bf n}_1,\,{\bf n}_2,\,{\bf n}_3,\,{\bf n}_4) =\,\cos\theta_1\,\cos\theta_2\,\cos\theta_3\,\cos\theta_4-\,\sin\theta_1\,
\sin\theta_2\,\sin\theta_3\,\sin\theta_4\,\cos\,\left(\,\phi_1+\phi_2-\phi_3-\phi_4\right). \label{q-preghz}
\end{equation}
For simplicity, the authors then restrict the directions to the $\mathrm{x}$-$\mathrm{y}$ plane, which simplifies the expectation function to
\begin{equation}
\left.{\cal E}^{\Psi_{\bf z}}_{{\!}_{QM}}({\bf n}_1,\,{\bf n}_2,\,{\bf n}_3,\,{\bf n}_4)\right|_{\text{x-y}}=\;-\,\cos\left(\,\phi_1\,+\,\phi_2\,-\,\phi_3\,-\,\phi_4\,\right). \label{followsfor}
\end{equation}
Next, the authors consider two special cases of perfect correlation, since that is one of the premisses of the argument by EPR. Accordingly, they set the sum of angles $\phi_1\,+\,\phi_2\,-\,\phi_3\,-\,\phi_4=0$, to obtain from (\ref{followsfor}) the expectation value 
\begin{equation}
\left.{\cal E}^{\Psi_{\bf z}}_{{\!}_{QM}}({\bf n}_1,\,{\bf n}_2,\,{\bf n}_3,\,{\bf n}_4)\right|_{\text{x-y}}=\;-1\,. \label{perfect1}
\end{equation}
Similarly, they set the sum of angles $\phi_1\,+\,\phi_2\,-\,\phi_3\,-\,\phi_4=\pi$ to obtain 
\begin{equation}
\left.{\cal E}^{\Psi_{\bf z}}_{{\!}_{QM}}({\bf n}_1,\,{\bf n}_2,\,{\bf n}_3,\,{\bf n}_4)\right|_{\text{x-y}}=\;+1\,. \label{perfect2}
\end{equation}
Thus, for these cases the expectation values remain constant for all runs of their thought experiment. Here, by perfect correlation the authors mean that, with four Stern-Gerlach analyzers set at angles satisfying the conditions (\ref{perfect1}) and (\ref{perfect2}) in their experimental setup shown in Figure~2 of \cite{GHZ}, knowledge of the outcomes for any three particles enables a prediction with certainty of the outcome for the fourth, in analogy with the two-particle case considered by Bell \cite{Bell-1964}.

Now, in analogy with the singlet state of Section~\ref{Sec-C}, let us analyze the local-realistic counterpart of the predictions of the state (\ref{ghz-single}) for the observables (\ref{fullop}). The eigenvalues of (\ref{fullop}), expressed in the dispersion-free state $\left|\,\Psi_{\bf z},\,\lambda\,\right)$, are 
\begin{equation}
\omega(c,\,\lambda) = {\mathscr A}{\mathscr B}{\mathscr C}{\mathscr D}({\bf n}_1,\,{\bf n}_2,\,{\bf n}_3,\,{\bf n}_4,\,\lambda)=\pm1, \label{eigen2}
\end{equation}
where ${\mathscr A}=\pm1$, ${\mathscr B}=\pm1$, ${\mathscr C}=\pm1$, and ${\mathscr D}=\pm1$ are the possible individual outcomes of spin measurements along the respective directions when the dispersion-free state of the four-particle system is $\left|\,\Psi_{\bf z},\,\lambda\,\right)$. In analogy with the locality\break condition (\ref{fact}), in the present case local causality demands that the above eigenvalues must be factorized as follows:
\begin{equation}
\omega(c,\,\lambda) = {\mathscr A}{\mathscr B}{\mathscr C}{\mathscr D}({\bf n}_1,\,{\bf n}_2,\,{\bf n}_3,\,{\bf n}_4,\,\lambda)={\mathscr A}({\bf n}_1\,,\,\lambda)\;{\mathscr B}({\bf n}_2\,,\,\lambda)\;{\mathscr C}({\bf n}_3\,,\,\lambda)\;{\mathscr D}({\bf n}_4\,,\,\lambda)=\pm1\,. \label{GHZlocality}
\end{equation}
Consequently, in the $\mathrm{x}$-$\mathrm{y}$ plane a local-realistic counterpart of the quantum mechanical expectation value (\ref{followsfor})  should~be
\begin{equation}
\left.{\cal E}^{\rm GHZ}_{{\!}_{LR}}({\bf n}_1,\,{\bf n}_2,\,{\bf n}_3,\,{\bf n}_4)\right|_{\text{x-y}}=\int_{\mathscr L}{\mathscr A}({\bf n}_1,\,\lambda)\,{\mathscr B}({\bf n}_2,\,\lambda)\,{\mathscr C}({\bf n}_3,\,\lambda)\,{\mathscr D}({\bf n}_4,\,\lambda)\;\,p(\lambda)\,d\lambda \,=\,-\,\cos\left(\,\phi_1\,+\,\phi_2\,-\,\phi_3\,-\,\phi_4\,\right).
\end{equation}
In other words, in analogy with (\ref{first}), it should be an ensemble average of (\ref{GHZlocality}) over the probability distribution $p(\lambda)$. However, for the case of perfect correlation the authors set $p(\lambda)=1$, which reduces the above expectation values to 
\begin{equation}
\left.{\cal E}^{\rm GHZ}_{{\!}_{LR}}({\bf n}_1,\,{\bf n}_2,\,{\bf n}_3,\,{\bf n}_4)\right|_{\text{x-y}}\,=\,{\mathscr A}({\bf n}_1,\,\lambda)\,{\mathscr B}({\bf n}_2,\,\lambda)\,{\mathscr C}({\bf n}_3,\,\lambda)\,{\mathscr D}({\bf n}_4,\,\lambda) \,=\,\pm1,
\end{equation}
with the local-realistic counterparts of the two quantum mechanical alternatives
(\ref{perfect1}) and (\ref{perfect2}) reducing to the following:
\begin{subequations}
\begin{equation}
\text{If}\;\,\phi_1\,+\,\phi_2\,-\,\phi_3\,-\,\phi_4=0,\;\text{then}\;\;\omega(c,\,\lambda) = {\mathscr A}({\bf n}_1,\,\lambda)\,{\mathscr B}({\bf n}_2,\,\lambda)\,{\mathscr C}({\bf n}_3,\,\lambda)\,{\mathscr D}({\bf n}_4,\,\lambda) \,=\,-1\, \label{67}
\end{equation}
and
\begin{equation}
\text{If}\;\,\phi_1\,+\,\phi_2\,-\,\phi_3\,-\,\phi_4=\pi,\;\text{then}\;\;\omega(c,\,\lambda) = {\mathscr A}({\bf n}_1,\,\lambda)\,{\mathscr B}({\bf n}_2,\,\lambda)\,{\mathscr C}({\bf n}_3,\,\lambda)\,{\mathscr D}({\bf n}_4,\,\lambda) \,=\,+1. \label{68}
\end{equation}
\end{subequations}
Next, they consider implications of (\ref{67}) for a set of four possible results specified by an azimuthal angle $\phi$ as follows:
\begin{subequations}
\begin{align}
&\omega(c',\,\lambda) = {\mathscr A}(0,\,\lambda)\,{\mathscr B}(0,\,\lambda)\,{\mathscr C}(0,\,\lambda)\,{\mathscr D}(0,\,\lambda)=-1, \label{69a} \\
&\omega(c'',\,\lambda) = {\mathscr A}(\phi,\,\lambda)\,{\mathscr B}(0,\,\lambda)\,{\mathscr C}(\phi,\,\lambda)\,{\mathscr D}(0,\,\lambda)=-1, \label{69b}\\
&\omega(c''',\,\lambda) = {\mathscr A}(\phi,\,\lambda)\,{\mathscr B}(0,\,\lambda)\,{\mathscr C}(0,\,\lambda)\,{\mathscr D}(\phi,\,\lambda)=-1, \label{69c} \\
\text{and}\;\;&\omega(c'''',\,\lambda) = {\mathscr A}(2\phi,\,\lambda)\,{\mathscr B}(0,\,\lambda)\,{\mathscr C}(\phi,\,\lambda)\,{\mathscr D}(\phi,\,\lambda)=-1. \label{69d}
\end{align}
\end{subequations}
Thus, if we define vectors ${\bf p}$ and ${\bf q}$ in terms of the angle $\phi$ and unit vectors ${\bf x}$ and ${\bf y}$ along axes $\mathrm{x}$ and $\mathrm{y}$, respectively,~as 
\begin{align}
{\bf p}&:= \cos(\phi)\,{\bf x}+\sin(\phi)\,{\bf y} \label{p} \\
\text{and}\;\;{\bf q}&:= \cos(2\phi)\,{\bf x}+\sin(2\phi)\,{\bf y},
\end{align}
then, in the experimental setup shown in Figure~2 of \cite{GHZ}, the results (\ref{69a}) to (\ref{69d}) of spin measurements supposed by the authors to occur would be along the following set of four alternative directions or contexts in physical space:
\begin{subequations}
\begin{align}
c' &= \{\,{\bf n_1'} = {\bf x},\;{\bf n_2'} = {\bf x},\;{\bf n_3'} = {\bf x},\;{\bf n_4'} = {\bf x}\,\}, \\
c'' &= \{\,{\bf n_1''} = {\bf p},\;{\bf n_2''} = {\bf x},\;{\bf n_3''} = {\bf p},\;{\bf n_4''} = {\bf x}\,\}, \\
c''' &= \{\,{\bf n_1'''} = {\bf p},\;{\bf n_2'''} = {\bf x},\;{\bf n_3'''} = {\bf x},\;{\bf n_4'''} = {\bf p}\,\}, \\
\text{and}\;\;c'''' &= \{\,{\bf n_1''''} = {\bf q},\;{\bf n_2''''} = {\bf x},\;{\bf n_3''''} = {\bf p},\;{\bf n_4''''} = {\bf p}\,\}.
\end{align}
\end{subequations}
Using (\ref{fullop}), the four quantum mechanical observables along the above set of alternatively possible directions are then  
\begin{subequations}
\begin{align}
\Omega(c')&=({\boldsymbol\sigma}\cdot{\bf x})_1\otimes
({\boldsymbol\sigma}\cdot{\bf x})_2\otimes({\boldsymbol\sigma}\cdot{\bf x})_3\otimes({\boldsymbol\sigma}\cdot{\bf x})_4\,, \label{73a} \\
\Omega(c'')&=({\boldsymbol\sigma}\cdot{\bf p})_1\otimes({\boldsymbol\sigma}\cdot{\bf x})_2\otimes({\boldsymbol\sigma}\cdot{\bf p})_3\otimes({\boldsymbol\sigma}\cdot{\bf x})_4\,, \label{73b} \\
\Omega(c''')&=({\boldsymbol\sigma}\cdot{\bf p})_1\otimes({\boldsymbol\sigma}\cdot{\bf x})_2\otimes({\boldsymbol\sigma}\cdot{\bf x})_3\otimes({\boldsymbol\sigma}\cdot{\bf p})_4\,, \label{73c} \\
\Omega(c'''')&=({\boldsymbol\sigma}\cdot{\bf q})_1\otimes({\boldsymbol\sigma}\cdot{\bf x})_2\otimes({\boldsymbol\sigma}\cdot{\bf p})_3\otimes({\boldsymbol\sigma}\cdot{\bf p})_4\,, \label{73d}
\end{align}
\end{subequations}
with their eigenvalues constrained to be (\ref{69a}) to (\ref{69d}), respectively, for the case (\ref{67}) of perfect correlation. Note that $\left[{\boldsymbol\sigma}\cdot{\bf x},\,{\boldsymbol\sigma}\cdot{\bf p}\right]=0$ if the operators ${\boldsymbol\sigma}\cdot{\bf x}$ and ${\boldsymbol\sigma}\cdot{\bf p}$ pertain to different particles within the same alternative or context. Consequently, the product rule (\ref{prod2}) holds within each alternative, and therefore equations (\ref{69a}) to (\ref{69d}) reproduced above from \cite{GHZ} are correct. On the other hand, $\left[{\boldsymbol\sigma}\cdot{\bf x},\,{\boldsymbol\sigma}\cdot{\bf p}\right]=2\,i\,\sin(\phi)\,{\boldsymbol\sigma}\cdot{\bf z}\not=0$ for $\phi\not=0$ (or $\not=$ a multiple of $\pi$)\break if ${\boldsymbol\sigma}\cdot{\bf x}$ and ${\boldsymbol\sigma}\cdot{\bf p}$ pertain to the same particle across different contexts or alternatives, and therefore $\left[\,\Omega(c'),\,\Omega(c'')\,\right]\not=0$, $\left[\,\Omega(c'),\,\Omega(c''')\,\right]\not=0$, and $\left[\,\Omega(c''),\,\Omega(c''')\,\right]\not=0$. As a result, for those cases, the product rule (\ref{productrule}) or (\ref{prod2}) does not hold:
\begin{equation}
\left(\,\Psi_{\bf z},\,\lambda\,|\,\Omega(c')\,\Omega(c'')\,\Omega(c''')\,|\,\Psi_{\bf z},\,\lambda\,\right) \,\not=\, \left(\,\Psi_{\bf z},\,\lambda\,|\,\Omega(c')\,|\,\Psi_{\bf z},\,\lambda\,\right)\,\left(\,\Psi_{\bf z},\,\lambda\,|\,\Omega(c'')\,|\,\Psi_{\bf z},\,\lambda\,\right)\,\left(\,\Psi_{\bf z},\,\lambda\,|\,\Omega(c''')\,|\,\Psi_{\bf z},\,\lambda\,\right). \label{notkjh}
\end{equation}
If we denote the eigenvalue of $\left\{\Omega(c')\,\Omega(c'')\,\Omega(c''')\right\}$ by $\widetilde{\omega}(\tilde{c},\,\lambda)$, then, using the ansatz (\ref{hidres}),  (\ref{notkjh}) can also be expressed~as
\begin{equation}
\widetilde{\omega}(\tilde{c},\,\lambda)\not=\,\omega(c',\,\lambda)\;\omega(c'',\,\lambda)\;\omega(c''',\,\lambda). 
\end{equation}
Thus, if the eigenvalues of non-commuting operators $\Omega(c')$, $\Omega(c'')$, and $\Omega(c''')$ are $\omega(c',\,\lambda)$, $\omega(c'',\,\lambda)$, and $\omega(c''',\,\lambda)$, respectively, then one cannot conclude that the eigenvalue of $\left\{\Omega(c')\,\Omega(c'')\,\Omega(c''')\right\}$ must be $\omega(c',\,\lambda)\,\omega(c'',\,\lambda)\,\omega(c''',\,\lambda)$.
As we will soon see, however, that even for these non-commuting observables pertaining to different alternatives or contexts, the authors of \cite{GHZ} illegitimately assumed a multiplicative expectation function for the dispersion-free states.

\subsubsection{Illegitimate use of multiplicative expectation functions and eigenvalues in \cite{GHZ}}

Returning to their derivation, in (\ref{68}) the authors use angles $\phi_1=\theta+\pi$, $\phi_2=0$, $\phi_3=\theta$, and $\phi_4=0$ to write it as
\begin{equation}
\omega(c''''',\,\lambda)={\mathscr A}(\theta+\pi,\,\lambda)\,{\mathscr B}(0,\,\lambda)\,{\mathscr C}(\theta,\,\lambda)\,{\mathscr D}(0,\,\lambda)=+1.
\end{equation}
The authors then compare this equality obtained from (\ref{68}) with the equality (\ref{69b}) obtained from (\ref{67}) to infer that
\begin{equation}
\omega(c''''',\,\lambda)={\mathscr A}(\theta+\pi,\,\lambda)\,{\mathscr B}(0,\,\lambda)\,{\mathscr C}(\theta,\,\lambda)\,{\mathscr D}(0,\,\lambda) \,=+1=-\,{\mathscr A}(\theta,\,\lambda)\,{\mathscr B}(0,\,\lambda)\,{\mathscr C}(\theta,\,\lambda)\,{\mathscr D}(0,\,\lambda)=-\,\omega(c'',\,\lambda),
\end{equation}
which immediately simplifies to
\begin{equation}
{\mathscr A}(\theta+\pi,\,\lambda) \,=-\,{\mathscr A}(\theta,\,\lambda). \label{73}
\end{equation}
This equation expresses perfect correlation and is consistent with the premisses of EPR. Thus, so far no mistake has occurred in the authors' reasoning or calculations, and neither has a contradiction with the premisses of EPR. But a serious and unrectifiable mistake occurs in the subsequent steps of their reasoning. To derive the alleged contradiction, the authors consider the products of the left- and right-hand sides of the equations (\ref{69a}), (\ref{69b}), and (\ref{69c}) to derive 
\begin{align}
\left(\,\Psi_{\bf z},\,\lambda\,|\,\Omega(c')\,\Omega(c'')\,\Omega(c''')\,|\,\Psi_{\bf z},\,\lambda\,\right) &= \left(\,\Psi_{\bf z},\,\lambda\,|\,\Omega(c')\,|\,\Psi_{\bf z},\,\lambda\,\right)\,\left(\,\Psi_{\bf z},\,\lambda\,|\,\Omega(c'')\,|\,\Psi_{\bf z},\,\lambda\,\right)\,\left(\,\Psi_{\bf z},\,\lambda\,|\,\Omega(c''')\,|\,\Psi_{\bf z},\,\lambda\,\right) \label{70} \\
\widetilde{\omega}(\tilde{c},\,\lambda)
&=\,\omega(c',\,\lambda)\;\omega(c'',\,\lambda)\;\omega(c''',\,\lambda) \label{71} \\
&=\{{\mathscr A}(0,\,\lambda)\,{\mathscr B}(0,\,\lambda)\,{\mathscr C}(0,\,\lambda)\,{\mathscr D}(0,\,\lambda)\}\times\{{\mathscr A}(\phi,\,\lambda)\,{\mathscr B}(0,\,\lambda)\,{\mathscr C}(\phi,\,\lambda)\,{\mathscr D}(0,\,\lambda)\} \notag \\
&\;\;\;\;\;\;\;\;\;\;\;\;\;\;\;\;\;\;\;\;\;\;\;\;\;\;\;\;\;\;\;\;\;\;\;\;\;\;\;\;\;\;\;\;\;\;\;\;\;\;\;\;\;\;\;\;\;\;\;\,\times\{{\mathscr A}(\phi,\,\lambda)\,{\mathscr B}(0,\,\lambda)\,{\mathscr C}(0,\,\lambda)\,{\mathscr D}(\phi,\,\lambda)\} \\
&= \{-1\}\times\{-1\}\times\{-1\} \\
&= -1 \label{74}
\end{align}
(see the endnote~15 in \cite{GHZ}, and also the sentence below Eq.~(24c) in \cite{GHZ}). Now, since each individual outcome such as ${\mathscr A}(\phi,\,\lambda)$ squares to $+1$ so that products ${\mathscr A}(\phi,\,\lambda){\mathscr A}(\phi,\,\lambda)=+1$, ${\mathscr C}(0,\,\lambda){\mathscr C}(0,\,\lambda)=+1$, {\it etc.}, this equality simplifies~to
\begin{equation}
\widetilde{\omega}(\tilde{c},\,\lambda)={\mathscr A}(0,\,\lambda)\,{\mathscr B}(0,\,\lambda)\,{\mathscr C}(\phi,\,\lambda)\,{\mathscr D}(\phi,\,\lambda) = -1. \label{soincorrect}
\end{equation}
The authors then compare the above equality with the equality (\ref{69d}) inferred earlier, to obtain the following relation:
\begin{equation}
\omega(c'''',\,\lambda)={\mathscr A}(2\phi,\,\lambda)\,{\mathscr B}(0,\,\lambda)\,{\mathscr C}(\phi,\,\lambda)\,{\mathscr D}(\phi,\,\lambda)
=-1={\mathscr A}(0,\,\lambda)\,{\mathscr B}(0,\,\lambda)\,{\mathscr C}(\phi,\,\lambda)\,{\mathscr D}(\phi,\,\lambda)=\,\widetilde{\omega}(\tilde{c},\,\lambda), \label{86}
\end{equation}
which immediately simplifies to
\begin{equation}
{\mathscr A}(2\phi,\,\lambda)=+\,{\mathscr A}(0,\,\lambda) =\,\text{constant for all}\;\phi. \label{75}
\end{equation}
The authors then compare this equality with the equality derived in (\ref{73}) by setting $\theta=0$ in (\ref{73}) and $\phi=\pi/2$ in (\ref{75}), leading to their alleged contradiction. The authors then claim: ``We have thus brought to the surface an inconsistency hidden in premisses (i)--(iv) [of EPR].'' However, this claim by the authors is not correct, because equation (\ref{75}) (which appears as equation (16) in \cite{GHZ}) is derived using the product rule (\ref{prod2}) illegitimately in steps (\ref{70}) to (\ref{74}), even though the observables (\ref{73a}) to (\ref{73c}) do not commute with each other \cite{Comment}. This mistake can be easily corrected as follows:

\subsubsection{Correction of the sign mistake in the derivation of equation (16) in \cite{GHZ}}

The incorrect step in the authors' derivation that leads them to the alleged inconsistency in the premisses of EPR is the inference (\ref{soincorrect}) above. That step is incorrect because --- as Bell stressed in his related criticism of von~Neumann's theorem \cite{Bell-1966} --- the result (\ref{soincorrect}) has been derived using products of eigenvalues of mutually non-commuting operators (\ref{73a}) to (\ref{73c}) that relate ``in a nontrivial way the results of experiments which cannot be performed simultaneously'' \cite{Bell-1966} (see also \cite{Grete}). Now, evidently, the quantity ${\mathscr A}(0,\,\lambda)\,{\mathscr B}(0,\,\lambda)\,{\mathscr C}(\phi,\,\lambda)\,{\mathscr D}(\phi,\,\lambda)$ in (\ref{soincorrect}) pertains to the combination of azimuthal angles adding up to $\phi_1\,+\,\phi_2\,-\,\phi_3\,-\,\phi_4=-2\phi$. However, this combination cannot satisfy the constraints (\ref{67}) of perfect correlation unless we set $\phi=0$, and in that case (\ref{soincorrect}), (\ref{69b}), (\ref{69c}), and (\ref{69d}) all reduce to (\ref{69a}). As a\break result, the equality (\ref{86}) reduces to a trivial identity and thus can no longer serve to derive the incorrect relation (\ref{75}). 

Alternatively, one may set $\phi=-\pi/2$ so that $\phi_1\,+\,\phi_2\,-\,\phi_3\,-\,\phi_4=-2\phi$ satisfies the constraint (\ref{68}) instead of (\ref{67}). But, in that case, (\ref{soincorrect}) must be changed to $\widetilde{\omega}(\tilde{c},\,\lambda)={\mathscr A}(0,\,\lambda)\,{\mathscr B}(0,\,\lambda)\,{\mathscr C}(\phi,\,\lambda)\,{\mathscr D}(\phi,\,\lambda)=+1$, reducing (\ref{86})~to
\begin{equation}
\omega(c'''',\,\lambda)={\mathscr A}(2\phi,\,\lambda)\,{\mathscr B}(0,\,\lambda)\,{\mathscr C}(\phi,\,\lambda)\,{\mathscr D}(\phi,\,\lambda)
=-1=-\,{\mathscr A}(0,\,\lambda)\,{\mathscr B}(0,\,\lambda)\,{\mathscr C}(\phi,\,\lambda)\,{\mathscr D}(\phi,\,\lambda)=\,\widetilde{\omega}(\tilde{c},\,\lambda). \label{86new}
\end{equation}
However, for $\phi=-\pi/2$ this equality simplifies to
\begin{equation}
{\mathscr A}(-\pi,\,\lambda)=-\,{\mathscr A}(0,\,\lambda), \label{75inter}
\end{equation}
which is just an expression of perfect correlation assumed by EPR, thereby neutralizing the claim of contradiction.

Thus, as long as the quantity  ${\mathscr A}(0,\,\lambda)\,{\mathscr B}(0,\,\lambda)\,{\mathscr C}(\phi,\,\lambda)\,{\mathscr D}(\phi,\,\lambda)$ is required to respect a perfect correlation constraint (\ref{67}) or (\ref{68}), no contradiction with the premisses of EPR can be derived. However, one may insist that this quantity in question serves as an intermediate step in the derivation of (\ref{75}) and therefore need not satisfy either of the two constraints (\ref{67}) or (\ref{68}). But it is evident that the quantity ${\mathscr A}(0,\,\lambda)\,{\mathscr  B}(0,\,\lambda)\,{\mathscr C}(\phi,\,\lambda)\,{\mathscr D}(\phi,\,\lambda)$ is a genuine eigenvalue of a possible observable in the GHZ system of four particles, albeit for a context ({\it i.e.}, a set of directions) different from the ones the authors have considered to derive contradiction. It is a {\it bona fide} eigenvalue of the following observable
\begin{equation}
\widetilde{\Omega}(\tilde{c})=({\boldsymbol\sigma}\cdot{\bf x})_1\otimes({\boldsymbol\sigma}\cdot{\bf x})_2\otimes({\boldsymbol\sigma}\cdot{\bf p})_3\otimes({\boldsymbol\sigma}\cdot{\bf p})_4\,. \label{difobs}
\end{equation}
The same is true of the quantity appearing in the authors' equation (15), namely ${\mathscr A}(2\phi,\,\lambda)\,{\mathscr  B}(0,\,\lambda)\,{\mathscr C}(0,\,\lambda)\,{\mathscr D}(0,\,\lambda)$, which the authors have derived indirectly via rather convoluted steps. It too is a {\it bona fide} eigenvalue of the observable
\begin{equation}
\overline{\Omega}(\overline{c\,})=({\boldsymbol\sigma}\cdot{\bf q})_1\otimes({\boldsymbol\sigma}\cdot{\bf x})_2\otimes({\boldsymbol\sigma}\cdot{\bf x})_3\otimes({\boldsymbol\sigma}\cdot{\bf x})_4\,. \label{difobs2}
\end{equation}
Now, by construction, all observables such as (\ref{fullop}) of the GHZ system can only take two eigenvalues: $+1$ or $-1$. On the other hand, the eigenvalue $\widetilde{\omega}(\tilde{c},\,\lambda)={\mathscr A}(0,\,\lambda)\,{\mathscr B}(0,\,\lambda)\,{\mathscr C}(\phi,\,\lambda)\,{\mathscr D}(\phi,\,\lambda)$ of the observable (\ref{difobs}) cannot be $-1$ because that value has been derived by employing the incorrect steps (\ref{70}) to (\ref{74}) involving illegitimate use of the product rule (\ref{prod2}). Those steps can only give an incorrect value for $\widetilde{\omega}(\tilde{c},\,\lambda)$ for non-commuting observables constituting the product. Therefore, since the only other possible eigenvalue of the observable (\ref{difobs}) within the scenario under consideration is $+1$, the correct eigenvalue of (\ref{difobs}) is necessarily equal to $+1$. Consequently, the correct version of (\ref{soincorrect}) is necessarily
\begin{equation}
\widetilde{\omega}(\tilde{c},\,\lambda)={\mathscr A}(0,\,\lambda)\,{\mathscr B}(0,\,\lambda)\,{\mathscr C}(\phi,\,\lambda)\,{\mathscr D}(\phi,\,\lambda) = +1. \label{socorrect}
\end{equation}
This equality can now be compared with the equality (\ref{69d}) as we did before for (\ref{soincorrect}), to obtain the following relation:
\begin{equation}
\omega(c'''',\,\lambda)={\mathscr A}(2\phi,\,\lambda)\,{\mathscr B}(0,\,\lambda)\,{\mathscr C}(\phi,\,\lambda)\,{\mathscr D}(\phi,\,\lambda)
=-1=-\,{\mathscr A}(0,\,\lambda)\,{\mathscr B}(0,\,\lambda)\,{\mathscr C}(\phi,\,\lambda)\,{\mathscr D}(\phi,\,\lambda)=\,\widetilde{\omega}(\tilde{c},\,\lambda),
\end{equation}
which immediately simplifies to
\begin{equation}
{\mathscr A}(2\phi,\,\lambda)=-\,{\mathscr A}(0,\,\lambda). \label{75new}
\end{equation}
As before, we now compare this equality with that in (\ref{73}) by setting $\theta=0$ in (\ref{73}) and $\phi=\pi/2$ in (\ref{75new}), which yields
\begin{equation}
{\mathscr A}(\pi,\,\lambda)=-\,{\mathscr A}(0,\,\lambda). \label{correctness}
\end{equation}
This equation expresses a perfect correlation consistent with the premisses of EPR, thus refuting the claim by GHZ.

\subsubsection{Proof of the consistency of the premisses of the EPR program by direct computation}

To remove any doubt about the correctness of (\ref{correctness}), we can also directly verify (\ref{socorrect}) that leads to it, by computing the\break product $\widetilde{\Omega}(\tilde{c})=\Omega(c')\,\Omega(c'')\,\Omega(c''')$ of the three operators involved in (\ref{socorrect}), using the definitions (\ref{73a}), (\ref{73b}), and (\ref{73c}):
\begin{align}
\widetilde{\Omega}(\tilde{c})=\Omega(c')\,\Omega(c'')\,\Omega(c''')&=\left\{({\boldsymbol\sigma}\cdot{\bf x})_1\otimes
({\boldsymbol\sigma}\cdot{\bf x})_2\otimes({\boldsymbol\sigma}\cdot{\bf x})_3\otimes({\boldsymbol\sigma}\cdot{\bf x})_4\right\} \notag \\
&\;\;\;\;\;\;\;\;\;\;\;\times\left\{({\boldsymbol\sigma}\cdot{\bf p})_1\otimes({\boldsymbol\sigma}\cdot{\bf x})_2\otimes({\boldsymbol\sigma}\cdot{\bf p})_3\otimes({\boldsymbol\sigma}\cdot{\bf x})_4\right\} \notag \\
&\;\;\;\;\;\;\;\;\;\;\;\;\;\;\;\;\;\;\;\;\times\left\{({\boldsymbol\sigma}\cdot{\bf p})_1\otimes({\boldsymbol\sigma}\cdot{\bf x})_2\otimes({\boldsymbol\sigma}\cdot{\bf x})_3\otimes({\boldsymbol\sigma}\cdot{\bf p})_4\right\}. \label{94n0}
\end{align}
Now, for the same particles, the identities $({\boldsymbol\sigma}\cdot{\bf x})^2=\dbl$ and $({\boldsymbol\sigma}\cdot{\bf p})^2=\dbl$ hold. Consequently, this product reduces to 
\begin{equation}
\widetilde{\Omega}(\tilde{c})=\Omega(c')\,\Omega(c'')\,\Omega(c''')=({\boldsymbol\sigma}\cdot{\bf x})_1\otimes
({\boldsymbol\sigma}\cdot{\bf x})_2\otimes\left[({\boldsymbol\sigma}\cdot{\bf x})({\boldsymbol\sigma}\cdot{\bf p})({\boldsymbol\sigma}\cdot{\bf x})\right]_3\otimes({\boldsymbol\sigma}\cdot{\bf p})_4\,. \label{3op}
\end{equation}
Next, using the definition (\ref{p}) of the direction ${\bf p}$ in Fig.~2 of \cite{GHZ}, we work out the operator $\left[({\boldsymbol\sigma}\cdot{\bf x})({\boldsymbol\sigma}\cdot{\bf p})({\boldsymbol\sigma}\cdot{\bf x})\right]_3$ as
\begin{align}
\left[({\boldsymbol\sigma}\cdot{\bf x})({\boldsymbol\sigma}\cdot{\bf p})({\boldsymbol\sigma}\cdot{\bf x})\right]_3
&= \left[({\boldsymbol\sigma}\cdot{\bf x})\,({\boldsymbol\sigma}\cdot\left\{\cos(\phi)\,{\bf x}+\sin(\phi)\,{\bf y}\right\})\,({\boldsymbol\sigma}\cdot{\bf x})\right]_3\\
&=\left[\cos(\phi)\,({\boldsymbol\sigma}\cdot{\bf x})^2({\boldsymbol\sigma}\cdot{\bf x})+\sin(\phi)\,({\boldsymbol\sigma}\cdot{\bf x})({\boldsymbol\sigma}\cdot{\bf y})({\boldsymbol\sigma}\cdot{\bf x})\right]_3 \\
&=\left[\cos(\phi)\,\dbl\,({\boldsymbol\sigma}\cdot{\bf x})-\,\sin(\phi)\,({\boldsymbol\sigma}\cdot{\bf y})\right]_3 \\
&=\left[{\boldsymbol\sigma}\cdot\left\{\cos(-\,\phi)\,{\bf x}+\sin(-\,\phi)\,{\bf y}\right\}\right]_3 \\
&=\left[{\boldsymbol\sigma}\cdot{\bf p}(-\,\phi)\right]_3, \label{101yes}
\end{align}
where I have used  $({\boldsymbol\sigma}\cdot{\bf x})^2=\dbl$ and the standard properties of Pauli matrices to derive $({\boldsymbol\sigma}\cdot{\bf x})({\boldsymbol\sigma}\cdot{\bf y})({\boldsymbol\sigma}\cdot{\bf x})=-\,({\boldsymbol\sigma}\cdot{\bf y})$. Using (\ref{101yes}), we can now express the product $\widetilde{\Omega}(\tilde{c})=\Omega(c')\,\Omega(c'')\,\Omega(c''')$ of three non-commuting operators in (\ref{94n0}) as
\begin{equation}
\widetilde{\Omega}(\tilde{c})=\Omega(c')\,\Omega(c'')\,\Omega(c''')=({\boldsymbol\sigma}\cdot{\bf x})_1\otimes
({\boldsymbol\sigma}\cdot{\bf x})_2\otimes\left[{\boldsymbol\sigma}\cdot{\bf p}(-\,\phi)\right]_3\otimes\left[{\boldsymbol\sigma}\cdot{\bf p}(+\,\phi)\right]_4. \label{evi101}
\end{equation}
Since the first perfect correlation condition $\phi_1\,+\,\phi_2\,-\,\phi_3\,-\,\phi_4=0$ from (\ref{67}) is now satisfied for this operator for any\break value of $\phi$ [because it is evident from (\ref{evi101}) that we now have $\phi_1=0$, $\phi_2=0$, $\phi_3=-\,\phi$, and $\phi_4=+\,\phi$], its eigenvalue~is 
\begin{equation}
{\mathscr A}(0,\,\lambda)\,{\mathscr B}(0,\,\lambda)\,{\mathscr C}(-\,\phi,\,\lambda)\,{\mathscr D}(+\,\phi,\,\lambda)=-1. \label{srect}
\end{equation}
But in the left-handed spherical coordinates with azimuthal angle $-\,\phi$, all signs of spin eigenvalues are reversed, giving
\begin{equation}
{\mathscr C}(-\,\phi,\,\lambda)=-\,{\mathscr C}(+\,\phi,\,\lambda).
\end{equation}
Consequently, for any azimuthal angle $\phi$ in the right-handed spherical coordinates used in \cite{GHZ}, (\ref{srect}) can be written~as
\begin{equation}
\widetilde{\omega}(\tilde{c},\,\lambda)={\mathscr A}(0,\,\lambda)\,{\mathscr B}(0,\,\lambda)\,{\mathscr C}(+\,\phi,\,\lambda)\,{\mathscr D}(+\,\phi,\,\lambda) = +1.
\end{equation}
We have thus verified the correctness of (\ref{socorrect}) that leads to (\ref{correctness}) by direct computation, independently of all previous arguments involving the product rule. This proves that the GHZ's result (\ref{soincorrect}) and their central claim (\ref{75}) are incorrect.

One may still object that, although the authors mention in their endnote~15, giving credit to Mermin, that the steps (\ref{71}) to (\ref{75}) I have used above lead more directly to (\ref{75}) [{\it i.e.}, to their equation (16)], and they also use similar steps in their equations (24a) to (24d) for the case without spin, their equations (13a) to (16) involve different manipulations that do not seem to rely on using the product rule (\ref{prod2}). In their equations (13a) to (16) the authors use convoluted steps involving {\it ad hoc} comparisons of eigenvalues of non-commuting observables with different contexts exploiting numerical coincidences, division of their equation (13a) by their equation (13b) to arrive at their equation (14a), and inversion of measurement results ${\mathscr D}(\phi,\,\lambda)=\pm1$ and ${\mathscr D}(0,\,\lambda)\pm1$. These strange manipulations not only do serious violence to the underlying physics of their thought experiment but also mathematically obfuscate what they are doing by obscuring the fact that what they are actually doing is using the product rule (\ref{prod2}) illegitimately for non-commuting observables to arrive at their key equation (16). By contrast, the suggestion by Mermin to the authors I noted above, which I have followed in steps (\ref{71}) to (\ref{75}) verbatim, makes their illegitimate use of the product rule quite transparent.

It is also worth noting that no separate analysis for the scenario without spin presented by the authors in their Section~IV is needed as it uses the same set of equations to derive an alleged contradiction, as the authors acknowledge.

\subsection{Hardy's variant of Bell's theorem is not ``a proof of non-locality''}\label{Hardy-flaw}

Contrary to its stated claim, Hardy's variant of Bell's theorem is not ``a proof of non-locality'' but an elementary demonstration of the Kochen-Specker theorem \cite{Bell-1966,Kochen}. As we noted in Section~\ref{Sec-B}, the Kochen-Specker theorem does not rule out contextual hidden variable theories; and in the previous sections I demonstrated that contextuality does not necessitate ``non-locality.'' In this section, I demonstrate that Hardy's variant merely rules out non-contextuality. To that end, in place of the notations used so far, I will mostly use Hardy's notations in \cite{Hardy} to avoid any confusion.

Hardy considers a specific two-particle entangled state formed in an experimental setup consisting of two overlapping Mach-Zehnder interferometers, one for positrons and another for electrons. The two interferometers are arranged so that electrons and positrons take particular separate paths, and subsequently meet and annihilate one another with probability equal to 1. The quantum state of the two particles can be written in the following four equivalent forms:
\begin{subequations}
\begin{align}
& |\Psi_{\mathrm{H}}\rangle=N\left(A B\left|u_{1}\right\rangle\left|v_{2}\right\rangle+A B\left|v_{1}\right\rangle\left|u_{2}\right\rangle+B^{2}\left|v_{1}\right\rangle\left|v_{2}\right\rangle\right) \label{13a} \\
& |\Psi_{\mathrm{H}}\rangle=N\left(\left|c_{1}\right\rangle\left(A\left|u_{2}\right\rangle+B\left|v_{2}\right\rangle\right)-A^{2}\left(A^{*}\left|c_{1}\right\rangle-B\left|d_{1}\right\rangle\right)\left|u_{2}\right\rangle\right) \label{13b} \\
& |\Psi_{\mathrm{H}}\rangle=N\left(\left(A\left|u_{1}\right\rangle+B\left|v_{1}\right\rangle\right)\left|c_{2}\right\rangle-A^{2}\left|u_{1}\right\rangle\left(A^{*}\left|c_{2}\right\rangle-B\left|d_{2}\right\rangle\right)\right) \label{13c} \\
& |\Psi_{\mathrm{H}}\rangle=N\left(\left|c_{1}\right\rangle\left|c_{2}\right\rangle-A^{2}\left(A^{*}\left|c_{1}\right\rangle-B\left|d_{1}\right\rangle\right)\left(A^{*}\left|c_{2}\right\rangle-B\left|d_{2}\right\rangle\right)\right), \label{13d}
\end{align}
\end{subequations}
where $\left|u_{i}\right\rangle$, $\left|v_{i}\right\rangle$, $\left|c_{i}\right\rangle$, and $\left|d_{i}\right\rangle$ are basis vectors in the two-particle Hilbert space; the indices $i=1$ and $2$ specify the\break particles $1$ and $2$, respectively; and $N$, $A$, and $B$ are complex numbers. Next, Hardy considers the projection operators
\begin{equation}
\widehat{U}_{i}=\left|u_{i}\right\rangle\left\langle u_{i}\right| \;\;\;\text{and}\;\;\; \widehat{D}_{j}=\left|d_{j}\right\rangle\left\langle d_{j}\right|,
\end{equation}
whose eigenvalues are $U_i=0$ or $1$ and $D_j=0$ or $1$, where $i$ and $j$ specify the particles $1$ or $2$. Note that $\widehat{U}_{i}$ and $\widehat{D}_{j}$ do not commute in general (which, as we will see, is the Achilles' heel of Hardy's argument). It is, therefore, not possible in general to measure $\widehat{U}_{i}$ and $\widehat{D}_{j}$ simultaneously on the same particle. Now, from (\ref{13a}) we see that if we measure $\widehat{U}_{1}$\break and $\widehat{U}_{2}$ on the particles $1$ and $2$ simultaneously (which is possible because $\widehat{U}_{1}$ and $\widehat{U}_{2}$ commute), then we would obtain
\begin{subequations}
\begin{equation}
U_{1} U_{2}=0, \label{14a}
\end{equation}
because there is no $\left|u_{1}\right\rangle\left|u_{2}\right\rangle$ term in (\ref{13a}). On the other hand, from (\ref{13b}) we see that if we measure $\widehat{D}_{1}$ on particle $1$ and $\widehat{U}_{2}$ on particle $2$, then we would obtain
\begin{equation}
U_{2}=1 \;\;\;\text{if}\;\;\; D_{1}=1, \label{14b}
\end{equation}
because only the $\left|d_{1}\right\rangle\left|u_{2}\right\rangle$ term in (\ref{13b}) contains $\left|d_{1}\right\rangle$. Similarly, from (\ref{13c}) we see that if we measure $\widehat{U}_{1}$ on particle $1$ and $\widehat{D}_{2}$ on particle $2$, then we would obtain
\begin{equation}
U_{1}=1 \;\;\;\text{if}\;\;\; D_{2}=1. \label{14c}
\end{equation}
Finally, from (\ref{13d}) we see that if we measure $\widehat{D}_{1}$ and $\widehat{D}_{2}$ simultaneously (which is possible because $\widehat{D}_{1}$ and $\widehat{D}_{2}$ also commute), then we would obtain
\begin{equation}
D_{1}=1 \;\;\text{and}\;\; D_{2}=1 \;\;\text{with probability}\;\; \left|N A^{2} B^{2}\right|^{2}. \label{14d}
\end{equation}
\end{subequations}

Using the quantum mechanical predictions (\ref{14a}) to (\ref{14d}), Hardy derives a seeming contradiction between the prediction (\ref{14a}) and its local-realistic counterpart dictated by any hidden variables $\lambda$. Note, however, that Hardy's derivation neglects to take into account the contexts of the measurement processes. By contrast, I reproduce Hardy's derivation below by explicitly including the contexts of all measurements involved. To that end, let us recall that, algebraically, the notion of contextuality is understood as follows \cite{Kochen}: If $\widehat{U}_{1}$ and $\widehat{U}_{2}$ are two commuting operators, then there always exists a non-degenerate operator $\widehat{C}_{U}$ and functions $f_1$ and $f_2$ such that $\widehat{U}_{1}=f_1(\widehat{C}_{U})$ and $\widehat{U}_{2}=f_2(\widehat{C}_{U})$. Evidently, in this case, it is possible to measure $\widehat{U}_{1}$ and $\widehat{U}_{2}$ simultaneously because it is only necessary to measure $\widehat{C}_{U}$ and apply the functions $f_1$ and $f_2$ to the measured value ${C}_{U}$ of $\widehat{C}_{U}$ to obtain the values of $\widehat{U}_{1}$ and $\widehat{U}_{2}$. It is therefore\break reasonable to use ${C}_{U}$ to denote the context of simultaneously possible measurements of $\widehat{U}_{1}$ and $\widehat{U}_{2}$. This mathematical definition of context algebraically encodes and environmentally identifies with the context of a chosen experimental setup in which all physical observables are mutually compatible. On the other hand, if $\widehat{D}_{1}$ and $\widehat{D}_{2}$ are another two commuting operators, then again there would exist a non-degenerate operator $\widehat{C}_{D}$ and functions $g_1$ and $g_2$ such that $\widehat{D}_{1}=g_1(\widehat{C}_{D})$ and $\widehat{D}_{2}=g_2(\widehat{C}_{D})$; but if $\widehat{U}_{i}$ and $\widehat{D}_{j}$ do not commute, then the contexts ${C}_{U}$ and ${C}_{D}$ cannot be the same:
\begin{equation}
{C}_{U}\not={C}_{D}\;\;\;\text{if}\;\;\;\left[\widehat{U}_{i},\,\widehat{D}_{j}\right]\not=0. 
\end{equation}
Consequently, comparisons of results are meaningful only within the same context, not across different contexts. The conceptual reasons behind this are the same as those recognized by Grete Hermann \cite{Grete}, as we discussed in Section~\ref{Sec-E}. 

Now, let us parallel Hardy's derivation in \cite{Hardy} within a hidden variable theory. From (\ref{13a}) we see that if we measure\break $\widehat{U}_{1}$ and $\widehat{U}_{2}$ on particles $1$ and $2$ simultaneously (which is possible because $\widehat{U}_{1}$ and $\widehat{U}_{2}$ commute), then we should obtain
\begin{equation}
U_{1}(C_U,\lambda)\,U_{2}(C_U,\lambda)=0, \label{14aCcon}
\end{equation}
because that is the quantum mechanical prediction (\ref{14a}) that we must reproduce in any hidden variable theory, where $C_U$ is the context of measurements. Now suppose that instead of $\widehat{U}_{1}$ and $\widehat{U}_{2}$ in the context $C_U$, the operators $\widehat{D}_{1}$ and $\widehat{D}_{2}$ in the context $C_D$ are measured and the results $D_{1}(C_D,\lambda)=1$ and $D_{2}(C_D,\lambda)=1$ are obtained for a specific $\lambda$ in a specific run of the experiment. That these results will be realized sometimes, follows from (\ref{14d}). Then, from the fact\break that we have $D_{1}(C_D,\lambda)=1$, it follows from (\ref{14b}) that if $\widehat{U}_{2}$ had been measured we would have obtained the result $U_{2}(C_D,\lambda)=1$. And, if we assume local causality, then we can say that, for this particular $\lambda$, we would have obtained $U_{2}(C_D,\lambda)=1$ even if $\widehat{U}_{1}$ had been measured on particle 1 instead of $\widehat{D}_{1}$, because according to the principle of local causality the choice of measurement on particle 1 cannot influence the outcome of any measurement on particle 2, and hence, for this run, $U_{2}(C_D,\lambda)$ must be determined by $\lambda$ to be equal to 1 --- {\it i.e.}, $U_{2}(C_D,\lambda)=1$. Similarly, we can deduce from the measurement result $D_{2}(C_D,\lambda)=1$ and (\ref{14c}) that $U_{1}(C_D,\lambda)=1$. Thus, for this run we would have 
\begin{equation}
U_{1}(C_D,\lambda)\,U_{2}(C_D,\lambda)=1. \label{14aDcon}
\end{equation}
Note, however, that, unlike Hardy's contradictory predictions $U_{1}(\lambda)\,U_{2}(\lambda)=0$ versus $U_{1}(\lambda)\,U_{2}(\lambda)=1$ that leads him to claim ``a proof of non-locality'', both left- and right-hand sides of our predictions (\ref{14aCcon}) and (\ref{14aDcon}) are different. Thus, we are not led to any contradiction. This is because we have not neglected to take into account the different contexts of measurements in the two cases. We have reasoned, following Hardy's reasoning, that if we had measured $\widehat{U}_{1}$ and $\widehat{U}_{2}$ in the context $C_D$ instead of $\widehat{D}_{1}$ and $\widehat{D}_{2}$ in the context $C_D$, then we would have obtained the result (\ref{14aDcon}) instead of the result (\ref{14aCcon}). However, in contrast to what Hardy concludes in \cite{Hardy}, our results (\ref{14aCcon}) and (\ref{14aDcon}) no longer contradict each other, because we have duly taken into account the correct contexts of measurements while counterfactually inferring the results $U_{1}(C_D,\lambda)=1$ and $U_{2}(C_D,\lambda)=1$ from the measurements of $\widehat{D}_{1}$ and $\widehat{D}_{2}$, because the latter measurements can only be performed within the context $C_D$. Consequently, unlike Hardy, we are not led to anything surprising about the fact that the values on the right-hand sides of (\ref{14aCcon}) and (\ref{14aDcon}) are different, because the contexts of measurements in those two cases are different, analogous to what Bell recognized in Section~5 of \cite{Bell-1966}. In other words, Hardy's claim of ``a proof of non-locality'' by demonstrating a contradiction between (\ref{14aCcon}) and (\ref{14aDcon}) is not correct. His argument in \cite{Hardy}\break is at best an example of the Kochen-Specker theorem \cite{Kochen,Bell-1966} that rules out non-contextual hidden variable theories.    

\subsection{Conclusion: Bell's theorem assumes its conclusion ({\it petitio principii})} \label{Sec-G}

Let me reiterate the main points discussed above.~Together, they demonstrate that Bell's theorem begs the question.

(1) The first point is that the derivation in Section~\ref{Sec-D} of the bounds of $\pm2$ on (\ref{combi}) for the dispersion-free counterpart $|\,\Psi,\,\lambda)$ of the singlet state (\ref{single}) must comply with the heuristics of the contextual hidden variable theories discussed in Section~\ref{Sec-B}.~Otherwise, the stringent bounds of $\pm2$ cannot be claimed to have any relevance for hidden variable theories. This requires compliance with the prescription (\ref{99}) that equates the quantum mechanical expectation values with their\break hidden variable counterparts for {\it all} observables, including any sums of observables, pertaining to the singlet system.

(2) The most charitable view of the equality (\ref{ladd}) is that it is an {\it assumption}, over and above those of locality, realism, and all other auxiliary assumptions required for deriving the inequalities (\ref{chsh}), because {\it it is valid only for commuting observables}. Far from being required by realism, it contradicts realism, because it fails to assign the correct eigenvalue (\ref{correct}) to the summed observable (\ref{op}) as its realistic counterpart, as required by the prescription (\ref{99}). Realism requires\break that all observables, including their sums, must be assigned unique eigenvalues, regardless of whether they are observed.

(3) Expectation values in dispersion-free states of hidden variable theories do not add linearly for observables that are not simultaneously measurable. And yet, Bell assumed linear additivity (\ref{ladd}) within a local hidden variable model. Conversely, in the light of the heuristics of contextual hidden variable theories we discussed in Section~\ref{Sec-B}, assuming (\ref{ladd})\break is equivalent to assuming that the spin observables ${\boldsymbol\sigma}_1\cdot{\bf a}\,\otimes\,{\boldsymbol\sigma}_2\cdot{\bf b}$, {\it etc.} commute with each other, but they do not.

(4) When the correct eigenvalue (\ref{correct}) is assigned to the summed operator (\ref{op}) replacing the incorrect step (\ref{ladd}), the bounds on Bell-CHSH sum (\ref{combi}) work out to be $\pm2\sqrt{2}$ instead of $\pm2$, thus mitigating the conclusion of Bell's theorem.

(5) As we proved in Section~\ref{Sec-D}, the assumption (\ref{ladd}) of the additivity of expectation values is equivalent to assuming the strong bounds of $\pm2$ on Bell-CHSH sum (\ref{combi}) of expectation values. In other words, (\ref{ladd}) and (\ref{chsh}) are tautologous.

(6) For observables that are not simultaneously measurable or commuting, the built-in linear additivity of integrals in step (\ref{ladd}) or (\ref{sumdisint}) leads to incorrect equality between averages of unequal physical quantities. Therefore, the view that this step is a harmless mathematical step is mistaken. It is, in fact, an unjustified assumption that is equivalent to the very thesis of the theorem to be proven, and is valid only in classical physics and/or for commuting observables.

The first four points above invalidate assumption (\ref{ladd}), and thus inequalities (\ref{chsh}) on physical grounds, and the last\break two demonstrate that Bell's theorem assumes its conclusion in a different guise, and is thus invalid on logical grounds.

Similar circularity in reasoning invalidates all variants of Bell's theorem. For example, as I explained in Section~\ref{GHZ-flaw}, the incorrect claim of the GHZ variant of Bell's theorem stems from a sign mistake in the equation (16) of \cite{GHZ}. But one may also view this mistake as stemming from circular reasoning. The authors implicitly assume a multiplicative expectation function as specified by the Product Rule (\ref{productrule}), even for the non-commuting observables (\ref{fullop}) involved in\break their thought experiment. They thereby assume their conclusion in a different guise in the premisses of their argument.

Bell's theorem is useful, however, for ruling out classical local theories. By relying on the assumption (\ref{ladd}), which is valid for classical theories \cite{Grete}, it proves that no classical local theory can reproduce all of the predictions of quantum\break mechanics. But no serious hidden variable theories I am aware of have ever advocated returning to classical~physics~\cite{Gudder}.

In this paper, I have focused on a formal or logical critique of Bell's theorem. Elsewhere \cite{RSOS,Local,RSOS-Reply}, I have developed a comprehensive local-realistic framework for understanding quantum correlations in terms of the geometry of the spatial part of one of the well-known solutions of Einstein's field equations of general relativity --- namely, that of a quaternionic 3-sphere --- taken as a physical space within which we are confined to perform Bell-test experiments. This\break framework is based on Clifford algebra and thus explicitly takes the non-commutativity of observables into account. It thus shows, constructively, that contextually local hidden variable theories are not ruled out by Bell-test experiments. Since, as we discussed in Section~\ref{Sec-C}, the formal proof of Bell's theorem is based on the entangled singlet state (\ref{single}), in \cite{Christian,IJTP,IEEE-1,IEEE-2,IEEE-3,IEEE-4,Symmetric} I have reproduced the correlations predicted by (\ref{single}) as a special case within the local-realistic framework proposed in \cite{RSOS,Local,RSOS-Reply}.~I especially recommend the calculations presented in \cite{IJTP} and \cite{Symmetric}, which also discuss a~macroscopic experiment that would be able to falsify the 3-sphere hypothesis I have proposed in these publications. Moreover, in \cite{RSOS,disproof-GHZ} I have reproduced exactly, not only the sinusoidal correlations predicted by quantum mechanics for the two-particle singlet state (\ref{single}), but also the prediction (\ref{q-preghz}) of the four-particle GHZ state (\ref{ghz-single}) and its three-particle analog, as well as all sixteen predictions of the Hardy state, as special cases in this local-realistic framework.

\appendix

\section*{Appendix}

\section{Dynamical equivalence of quantum mechanical description and Einstein's description} \label{D}

To establish the equivalence between two dynamics exhibited in Eq.~(\ref{Eh}), let us begin with the Ehrenfest equation,
\begin{equation}
\frac{d\,}{dt}\left\langle\,\psi\,|\,\Omega\,|\,\psi\,\right\rangle \,=\, \frac{1}{i\hbar}\left\langle\,\psi\,|\,\left[\,\Omega,\,H\,\right]\,|\,\psi\,\right\rangle \,+\, \langle\,\psi\,|\,\frac{\partial\Omega}{\partial t}\,|\,\psi\,\rangle, \label{eheh}
\end{equation}
where $H$ is a Hamiltonian operator.~Using prescription (\ref{77}) for the complete description, this equation can be written~as 
\begin{equation}
\frac{d\,}{dt}\int_{\mathscr L}
\left(\,\psi,\,\lambda\,|\,\Omega\,|\,\psi,\,\lambda\,\right) \,p(\lambda)\,d\lambda \,=\, \frac{1}{i\hbar}\int_{\mathscr L}
\left(\,\psi,\,\lambda\,|\,\left[\,\Omega,\,H\,\right]\,|\,\psi,\,\lambda\,\right) \,p(\lambda)\,d\lambda \,+
\int_{\mathscr L}
(\,\psi,\,\lambda\,|\,\frac{\partial\Omega}{\partial t}\,|\,\psi,\,\lambda\,) \;p(\lambda)\,d\lambda\,. \label{hami}
\end{equation}
Since in the ansatz (\ref{hidres}) we have implicitly assumed that the hidden variables $\lambda$ do not depend on time explicitly, the Leibniz integral rule for differentiation allows us to reduce the first term on the left-hand side of the above equation~to
\begin{equation}
\frac{d\,}{dt}\int_{\mathscr L}
\left(\,\psi,\,\lambda\,|\,\Omega\,|\,\psi,\,\lambda\,\right) \,p(\lambda)\,d\lambda \,= \frac{d\,}{dt}\int_{\mathscr L} \omega(\lambda)\;p(\lambda)\,d\lambda \,= \int_{\mathscr L} \left[\frac{d\,}{dt}\,\omega(\lambda)\right] p(\lambda)\,d\lambda\,. \label{aaa3}
\end{equation}
Next, we recall the Dirac rule for canonical quantization, which promotes the Poisson brackets to operators and sets
\begin{equation}
\left[\,\Omega,\,H\,\right] = i\hbar\,\widehat{\{\omega,\,{\cal H}\}}, \label{canoni}
\end{equation}
where $\omega$ is an eigenvalue of the observable $\Omega$ and the Hamiltonian function ${\cal H}$ is an eigenvalue of the corresponding operator $H$.~Then, the Poisson bracket $\{\omega,\,{\cal H}\}$ is an eigenvalue of the operator $\widehat{\{\omega,\,{\cal H}\}}$. Now it is well known that the Dirac rule is not unique in general when promoting classical variables to quantum operators. It is unique, however, for the reverse process of reducing the quantum operators to classical variables.~Given any two operators $\Omega_1$ and $\Omega_2$ with eigenvalues $\omega_1$ and $\omega_2$, their commutator $[\Omega_1,\,\Omega_2]$ uniquely reduces to the Poisson bracket $\{\omega_1,\,\omega_2\}$ in correspondence limit. Consequently, using (\ref{canoni}) and the ansatz (\ref{hidres}), we can transform the first term on the right-hand side of (\ref{hami}) to
\begin{equation}
\frac{1}{i\hbar}\int_{\mathscr L}
\left(\,\psi,\,\lambda\,|\,\left[\,\Omega,\,H\,\right]\,|\,\psi,\,\lambda\,\right) \,p(\lambda)\,d\lambda \,=\int_{\mathscr L}
(\,\psi,\,\lambda\,|\,\widehat{\{\omega,\,{\cal H}\}}\,|\,\psi,\,\lambda\,) \,p(\lambda)\,d\lambda \,=\int_{\mathscr L}\{\omega(\lambda),\,{\cal H}(\lambda)\}\;p(\lambda)\,d\lambda\,, \label{ssss}
\end{equation}
where a consistent application of ansatz (\ref{hidres}) to all observables of the system requires that $\{\omega,\,{\cal H}\}(\lambda)=\{\omega(\lambda),\,{\cal H}(\lambda)\}$.
Finally, using the ansatz (\ref{hidres}) for evaluating the second term on the right-hand side of (\ref{hami}), it is easy to prove that
\begin{equation}
\int_{\mathscr L}(\,\psi,\,\lambda\,|\,\frac{\partial\Omega}{\partial t}\,|\,\psi,\,\lambda\,) \,p(\lambda)\,d\lambda \,=\int_{\mathscr L}\frac{\partial\omega(\lambda)}{\partial t}\;p(\lambda)\,d\lambda\,. \label{hhh}
\end{equation}
Collecting all of the results obtained in (\ref{hami}), (\ref{aaa3}),(\ref{ssss}), and (\ref{hhh}) together, we can finally conclude using (\ref{eheh}) that
\begin{equation}
\left[\frac{d\,}{dt}\left\langle\,\psi\,|\,\Omega\,|\,\psi\,\right\rangle = \frac{1}{i\hbar}\left\langle\,\psi\,|\,\left[\,\Omega,\,H\,\right]\,|\,\psi\,\right\rangle + \langle\,\psi\,|\,\frac{\partial\Omega}{\partial t}\,|\,\psi\,\rangle\right] = \int_{\mathscr L}\left[\frac{d\;}{dt}\,\omega(\lambda) = \{\omega(\lambda),\,{\cal H}(\lambda)\} + \frac{\partial\,\omega(\lambda)}{\partial t}\right] p(\lambda)\,d\lambda\,. \label{Ehap}
\end{equation}
For specific values of $\lambda$ with $p(\lambda)=1$, the right-hand side of (\ref{Ehap}) reduces to classical equations of motion.~If, however, only a probability distribution $p(\lambda)$ of $\lambda$ is known, then Ehrenfest's equation in quantum mechanics can be understood as an average of an ensemble of classical dynamics, with each of its members specifying a trajectory in phase space.~All probabilistic statements about the system would then reduce to the incompleteness of our knowledge about the system.

\section{Demonstration of linearity of the expectation function $\left(\,\psi,\,\lambda\,|\,\Omega(c)\,|\,\psi,\,\lambda\,\right)$ assumed in (\ref{ladd})} \label{C}

For the demonstration of linearity of expectation function $\left(\,\psi,\,\lambda\,|\,\Omega(c)\,|\,\psi,\,\lambda\,\right)$ assumed in (\ref{ladd}), let us follow similar demonstration by Bohm and Bub in \cite{BohmBub} within the context of von Neumann's theorem. To that end, recall that in the Hilbert space model of quantum mechanics \cite{vonNeumann}, every ensemble can be characterized by a statistical operator, say ${\cal W(\psi)}$, of unit trace, representing the quantum state $|\,\psi\,\rangle$ of the system. In other words, there exists a linear Hermitian matrix ${\cal W}_{pq}(\psi)$ such that the expectation value in the state $|\,\psi\,\rangle$ of observable $\Omega$ can be expressed in the linear form
\begin{equation}
\left\langle\,\psi\,|\,\Omega\,|\,\psi\,\right\rangle =\text{Tr} \left\{{\cal W}(\psi)\,\Omega\right\} = \sum_{pq}\,{\cal W}_{qp}(\psi)\,\Omega_{pq}\,. \label{vonlinear}
\end{equation}
However, if dispersion in the measured values of $\Omega$ is assumed to be due to a distribution $p(\lambda)$ in the values of hidden variables $\lambda$ over the ensemble of the system as in (\ref{77}), then the expectation value of $\Omega$ should be the average over $\lambda$,
\begin{equation}
\left\langle\,\psi\,|\,\Omega\,|\,\psi\,\right\rangle \,=\int_{\mathscr L}\left(\,\psi,\,\lambda\,|\,\Omega\,|\,\psi,\,\lambda\,\right)\;p(\lambda)\,d\lambda = \sum_{pq}\int_{\mathscr L}{\cal W}_{qp}(\psi, \lambda)\,\Omega_{pq}\;p(\lambda)\,d\lambda\,, \label{speclam}
\end{equation}
which amounts to replacing the matrix ${\cal W}_{qp}(\psi)$ with its average $\overline{\cal W}_{qp}(\psi)$ over the distribution of hidden variables $\lambda$:
\begin{equation}
{\cal W}_{qp}(\psi) \longrightarrow \overline{\cal W}_{qp}(\psi) \,=  \int_{\mathscr L}{\cal W}_{qp}(\psi,\lambda)\;p(\lambda)\,d\lambda\,.
\end{equation}
For each particular value of $\lambda$, (\ref{speclam}) thus assumes expectation function to be of a linear form, similar to that in (\ref{vonlinear}): 
\begin{equation}
\left(\,\psi,\,\lambda\,|\,\Omega\,|\,\psi,\,\lambda\,\right) = \text{Tr} \left\{{\cal W}(\psi,\lambda)\,\Omega\right\} =  \sum_{pq}\,{\cal W}_{qp}(\psi, \lambda)\,\Omega_{pq}\,. \label{linof}
\end{equation}
By now it is well-recognized that demanding this linear form of expectation function $\left(\,\psi,\,\lambda\,|\,\Omega\,|\,\psi,\,\lambda\,\right)$ is not justifiable for hidden variable theories \cite{Bell-1966,Grete,Mermin,BohmBub}. Ironically, however, after criticizing von Neumann for retaining the linear form (\ref{linof}), Bell also ended up making the same mistake for contextual hidden variable theories, by implicitly assuming
\begin{equation}
\int_{\mathscr L}\left(\,\psi,\,\lambda\,|\,\Omega(c)\,|\,\psi,\,\lambda\,\right)\;p(\lambda)\,d\lambda  \,= \sum_{pq}\int_{\mathscr L}{\cal W}_{qp}(\psi, \lambda)\,\Omega_{pq}(c)\;p(\lambda)\,d\lambda
\end{equation}
in the equality (\ref{ladd}). It is not difficult to see that (\ref{ladd}) follows from the linearity (\ref{linof}) of the function $\left(\,\psi,\,\lambda\,|\,\Omega(c)\,|\,\psi,\,\lambda\,\right)$:
\begin{align}
\int_{\mathscr L}\left(\,\psi,\,\lambda\,|\,\Omega_1(c_1)+\Omega_2(c_2)\,|\,\psi,\,\lambda\,\right)\;p(\lambda)\,d\lambda \,&= \sum_{pq} \int_{\mathscr L}
{\cal W}_{qp}(\psi,\lambda)\,\left\{\,\Omega_1(c_1)+\Omega_2(c_2)\right\}_{pq}\;p(\lambda)\,d\lambda \\
\int_{\mathscr L}\widetilde{\omega}(\tilde{c},\lambda)\;p(\lambda)\,d\lambda \,&= \sum_{pq} \int_{\mathscr L}
{\cal W}_{qp}(\psi,\lambda)\,\left\{(\Omega_1)_{pq}(c_1)+(\Omega_2)_{pq}(c_2)\right\}\;p(\lambda)\,d\lambda \\
\int_{\mathscr L}\widetilde{\omega}(\tilde{c},\lambda)\;p(\lambda)\,d\lambda \,&= \sum_{pq} \int_{\mathscr L}
{\cal W}_{qp}(\psi,\lambda)\,(\Omega_1)_{pq}(c_1)\;p(\lambda)\,d\lambda \notag \\
\,&\quad\quad\quad\quad\quad\quad\quad\quad\quad + \sum_{pq} \int_{\mathscr L}
{\cal W}_{qp}(\psi,\lambda)\,(\Omega_2)_{pq}(c_2)\;p(\lambda)\,d\lambda \\
\int_{\mathscr L}\widetilde{\omega}(\tilde{c},\lambda)\;p(\lambda)\,d\lambda \,&=\int_{\mathscr L}\left(\,\psi,\,\lambda\,|\,\Omega_1(c_1)\,|\,\psi,\,\lambda\,\right) \;p(\lambda)\,d\lambda \;+ \int_{\mathscr L}\left(\,\psi,\,\lambda\,|\,\Omega_2(c_2)\,|\,\psi,\,\lambda\,\right)\;p(\lambda)\,d\lambda \\
\int_{\mathscr L}\widetilde{\omega}(\tilde{c},\lambda)\;p(\lambda)\,d\lambda \,&= \int_{\mathscr L}\omega_1(c_1,\lambda)\;p(\lambda)\,d\lambda \;+\int_{\mathscr L} \omega_2(c_2,\lambda)\;p(\lambda)\,d\lambda\,, \label{bothA10}
\end{align}
where only two terms of (\ref{ladd}) is used for simplicity, and the ansatz (\ref{hidres}) is used to give $\left(\,\psi,\,\lambda\,|\,\Omega_1(c_1)\,|\,\psi,\,\lambda\,\right)=\omega_1(c_1,\lambda)$, $\left(\,\psi,\,\lambda\,|\,\Omega_2(c_2)\,|\,\psi,\,\lambda\,\right)=\omega_2(c_2,\lambda)$, and $\left(\,\psi,\,\lambda\,|\,\Omega_1(c_1)+\Omega_2(c_2)\,|\,\psi,\,\lambda\,\right)=\widetilde{\omega}(\tilde{c},\lambda)$. Then linearity of anti-derivatives gives
\begin{equation}
\int_{\mathscr L}\widetilde{\omega}(\tilde{c},\lambda)\;p(\lambda)\,d\lambda \,= \int_{\mathscr L}\left\{\,\omega_1(c_1,\lambda) + \omega_2(c_2,\lambda)\right\}\;p(\lambda)\,d\lambda\,,
\end{equation}
implying that the eigenvalue $\widetilde{\omega}(\tilde{c},\lambda)$ of the sum $\Omega_1(c_1)+\Omega_2(c_2)$ of operators is equal to the sum of their eigenvalues:
\begin{equation}
\widetilde{\omega}(\tilde{c},\lambda) = \omega_1(c_1,\lambda) + \,\omega_2(c_2,\lambda)\,, \label{hjom}
\end{equation}
for {\it each} particular value of the hidden variables $\lambda$. But, as we noted in the last paragraph of Section~\ref{Sec-B}, this equality among the eigenvalues is mistaken in general (recall the example $\sqrt{2}\not=1+1$). It is valid only in classical physics, and for commuting observables. We have arrived at it by assuming the {\it linear} form of the expectation function $\left(\,\psi,\,\lambda\,|\,\Omega\,|\,\psi,\,\lambda\,\right)$\break defined in (\ref{linof}). Recognizing this difficulty, Bohm and Bub considered a more general, non-linear expectation function:
\begin{equation}
\left(\,\psi,\,\lambda\,|\,\Omega(c)\,|\,\psi,\,\lambda\,\right) = F(\psi,\lambda,\Omega_{pq}(c))\not=\sum_{pq}{\cal W}_{qp}(\psi, \lambda)\,\Omega_{pq}(c)\,, 
\end{equation}
where (at least for non-commuting observables) $F(\psi,\lambda,\Omega_{pq}(c))$ is a non-linear function of $\psi$, $\lambda$, and $\Omega_{pq}(c)$ such that
\begin{align}
\widetilde{\omega}(\tilde{c},\lambda) \not= \omega_1(c_1,\lambda) + \omega_2(c_2,\lambda)\,.
\end{align}
Indeed, there is no reason why these eigenvalues should not be determined by some non-linear $F(\psi,\lambda,\Omega_{pq}(c))$, giving
\begin{equation}
\int_{\mathscr L}\omega_1(c_1,\lambda)\;p(\lambda)\,d\lambda \;+\int_{\mathscr L} \omega_2(c_2,\lambda)\;p(\lambda)\,d\lambda \;= \int_{\mathscr L}\widetilde{\omega}(\tilde{c},\lambda)\;p(\lambda)\,d\lambda\;\not= \int_{\mathscr L}\left\{\,\omega_1(c_1,\lambda) + \omega_2(c_2,\lambda)\right\}\;p(\lambda)\,d\lambda\,.
\end{equation}
As Bell stressed in \cite{Bell-1966}, there is no reason for demanding (\ref{hjom}) of the dispersion-free state $|\,\psi,\,\lambda\,)$ for each value of ${\lambda}$.

\section{Separating the commuting and non-commuting parts of the summed operator (\ref{op})} \label{A}

Before considering the specific operator (\ref{op}),
in this appendix let us prove that, in general, the eigenvalue of a sum ${r\,{\mathcal R} + s\,{\mathcal S} + t\,{\mathcal T} + u\;{\mathcal U}}$ of operators is not equal to the sum $r\,{\mathscr R} + s\,{\mathscr S} + t\,{\mathscr T} + u\,{\mathscr U}$ of the individual eigenvalues of the operators ${\mathcal R}$, ${\mathcal S}$, ${\mathcal T}$, and ${\mathcal U}$, unless these operators commute with each other. Here $r$, $s$, $t$, and $u$ are real numbers. It is not difficult to prove this known fact by evaluating the square of the operator ${\{r\,{\mathcal R} + s\,{\mathcal S} + t\,{\mathcal T} + u\,{\mathcal U}\}}$ as follows:
\begin{align}
\{r\,{\mathcal R} + s\,{\mathcal S} + t\,{\mathcal T} + u\,{\mathcal U}\}\{r\,{\mathcal R} + s\,{\mathcal S} + t\,{\mathcal T} + u\,{\mathcal U}\} &= r^2{\mathcal R}^2 + rs\,{\mathcal R}{\mathcal S}+ rt\,{\mathcal R}{\mathcal T}+ ru\,{\mathcal R}{\mathcal U} \notag \\
&\;\;\;\;\;\;\;\;\;\;+ sr\,{\mathcal S}{\mathcal R}+ s^2 {\mathcal S}^2+ st\,{\mathcal S}{\mathcal T}+ su\,{\mathcal S}{\mathcal U} \notag \\
&\;\;\;\;\;\;\;\;\;\;+ tr\,{\mathcal T}{\mathcal R}+ ts\,{\mathcal T}{\mathcal S}+ t^2 {\mathcal T}^2+ tu\,{\mathcal T}{\mathcal U} \notag \\
&\;\;\;\;\;\;\;\;\;\;+ ur\,{\mathcal U}{\mathcal R}+ us\,{\mathcal U}{\mathcal S}+ ut\,{\mathcal U}{\mathcal T}+ u^2{\mathcal U}^2. \label{square}
\end{align}
Now, assuming that the operators ${\mathcal R}$, ${\mathcal S}$, ${\mathcal T}$, and ${\mathcal U}$ do not commute in general, let us define the following operators:
\begin{align}
{\mathcal L}&:={\mathcal S}{\mathcal R}-{\mathcal R}{\mathcal S} \;\Longleftrightarrow\; {\mathcal S}{\mathcal R} =  {\mathcal R}{\mathcal S}+{\mathcal L}, \label {def-i}\\
{\mathcal M}&:={\mathcal T}{\mathcal R}-{\mathcal R}{\mathcal T} \;\Longleftrightarrow\; {\mathcal T}{\mathcal R} =  {\mathcal R}{\mathcal T}+{\mathcal M}, \\
{\mathcal N}&:={\mathcal T}{\mathcal S}-{\mathcal S}{\mathcal T} \;\Longleftrightarrow\; {\mathcal T}{\mathcal S} =  {\mathcal S}{\mathcal T}+{\mathcal N}, \\
{\mathcal O}&:={\mathcal U}{\mathcal R}-{\mathcal R}{\mathcal U} \;\Longleftrightarrow\; {\mathcal U}{\mathcal R} =  {\mathcal R}{\mathcal U}+{\mathcal O}, \\
{\mathcal P}&:={\mathcal U}{\mathcal T}-{\mathcal T}{\mathcal U} \;\Longleftrightarrow\; {\mathcal U}{\mathcal T} =  {\mathcal T}{\mathcal U}+{\mathcal P}, \\
\text{and}\;\;\;{\mathcal Q}&:={\mathcal U}{\mathcal S}-{\mathcal S}{\mathcal U} \;\Longleftrightarrow\; {\mathcal U}{\mathcal S} =  {\mathcal S}{\mathcal U}+{\mathcal Q}. \label{def-f}
\end{align}
These operators would be null operators with vanishing eigenvalues if the operators ${\mathcal R}$, ${\mathcal S}$, ${\mathcal T}$, and ${\mathcal U}$ did commute with each other.~Using these relations for the operators ${\mathcal S}{\mathcal R}$, ${\mathcal T}{\mathcal R}$, ${\mathcal T}{\mathcal S}$, ${\mathcal U}{\mathcal R}$, ${\mathcal U}{\mathcal T}$ and ${\mathcal U}{\mathcal S}$, equation (\ref{square}) can be simplified~to
\begin{align}
\{r\,{\mathcal R} + s\,{\mathcal S} + t\,{\mathcal T} + u\,{\mathcal U}\}\{r\,{\mathcal R} + s\,{\mathcal S} + t\,{\mathcal T} + u\,{\mathcal U}\} &= r^2{\mathcal R}^2 + 2rs\,{\mathcal R}{\mathcal S}+ 2rt\,{\mathcal R}{\mathcal T}+ 2ru\,{\mathcal R}{\mathcal U} \notag \\
&\;\;\;\;\;\;\;\;\;\;+ rs\,{\mathcal L}+ s^2 {\mathcal S}^2+ 2st\,{\mathcal S}{\mathcal T}+ 2su\,{\mathcal S}{\mathcal U} \notag \\
&\;\;\;\;\;\;\;\;\;\;+ rt\,{\mathcal M}+ st\,{\mathcal N}+ t^2 {\mathcal T}^2+ 2tu\,{\mathcal T}{\mathcal U} \notag \\
&\;\;\;\;\;\;\;\;\;\;+ ru\,{\mathcal O}+ su\,{\mathcal Q}+ tu\,{\mathcal P}+ u^2 {\mathcal U}^2 \\
&=\{r\,{\mathcal R} + s\,{\mathcal S} + t\,{\mathcal T} + u\,{\mathcal U}\}_{\bf c}^2\,+\,{\mathcal Y}, \label{A9}
\end{align}
where
\begin{equation}
{\mathcal Y}:=rs\,{\mathcal L} + rt\,{\mathcal M} + st\,{\mathcal N}+ ru\,{\mathcal O}+ tu\,{\mathcal P} + su\,{\mathcal Q}\,. \label{line}
\end{equation}
We have thus separated out the commuting part ${\{r\,{\mathcal R} + s\,{\mathcal S} + t\,{\mathcal T} + u\,{\mathcal U}\}_{\bf c}}$ and the non-commuting part ${\mathcal Y}$ of the summed operator ${{\mathcal X}:=\{r\,{\mathcal R} + s\,{\mathcal S} + t\,{\mathcal T} + u\,{\mathcal U}\}}$. Note that the operators ${\mathcal L}$, ${\mathcal M}$, ${\mathcal N}$, ${\mathcal O}$, ${\mathcal P}$, and ${\mathcal Q}$ defined in (\ref{def-i}) to (\ref{def-f}) will not commute with each other in general unless their constituents ${\mathcal R}$, ${\mathcal S}$, ${\mathcal T}$, and ${\mathcal U}$ themselves are commuting. Next, we work out the eigenvalue ${\mathscr X}$ of the operator ${\mathcal X}$ in a normalized eigenstate $|\,\xi\,\rangle$ using the eigenvalue equations
\begin{equation}
{\mathcal X}\,|\,\xi\,\rangle = {\mathscr X}\,|\,\xi\,\rangle 
\end{equation}
and
\begin{equation}
{\mathcal X}\,{\mathcal X}\,|\,\xi\,\rangle =
{\mathcal X}\big\{{\mathcal X}\,|\,\xi\,\rangle\big\} = {\mathcal X}\,\big\{{\mathscr X}\,|\,\xi\,\rangle\big\} =
{\mathscr X}\,\big\{{\mathcal X}\,|\,\xi\,\rangle\big\} =
{\mathscr X}^2\,|\,\xi\,\rangle, \label{A11}
\end{equation}
in terms of the eigenvalues ${\mathscr R}$, ${\mathscr S}$, ${\mathscr T}$, and ${\mathscr U}$ of the operators ${\mathcal R}$, ${\mathcal S}$, ${\mathcal T}$, and ${\mathcal U}$ and the expectation value $\langle\,\xi\,|\,{\mathcal Y}\,|\,\xi\,\rangle$:
\begin{equation}
{\mathscr X} = \pm\sqrt{\langle\,\xi\,|\,{\mathcal X}\,{\mathcal X}\,|\,\xi\,\rangle}=\pm\sqrt{\langle\,\xi\,|\big\{r\,{\mathcal R} + s\,{\mathcal S} + t\,{\mathcal T} + u\,{\mathcal U}\big\}_{\bf c}^2\,|\,\xi\,\rangle+\langle\,\xi\,|\,{\mathcal Y}\,|\,\xi\,\rangle\,},
\end{equation}
where we have used (\ref{A9}). But the eigenvalue of the commuting part ${\{r\,{\mathcal R} + s\,{\mathcal S} + t\,{\mathcal T} + u\,{\mathcal U}\}_{\bf c}}$ of ${\mathcal X}$ is simply the linear sum ${r\,{\mathscr R} + s\,{\mathscr S} + t\,{\mathscr T} + u\,{\mathscr U}}$ of the eigenvalues of the operators ${\mathcal R}$, ${\mathcal S}$, ${\mathcal T}$, and ${\mathcal U}$. Consequently, using the equation analogous to (\ref{A11}) for the square of the operator ${\big\{r\,{\mathcal R} + s\,{\mathcal S} + t\,{\mathcal T} + u\,{\mathcal U}\big\}_{\bf c}}$ we can express the eigenvalue ${\mathscr X}$ of ${\mathcal X}$ as 
\begin{equation}
{\mathscr X}=\pm\sqrt{\big\{r\,{\mathscr R} + s\,{\mathscr S} + t\,{\mathscr T} + u\,{\mathscr U}\big\}^2 + \,\langle\,\xi\,|\,{\mathcal Y}\,|\,\xi\,\rangle\,}.
\label{A12}
\end{equation}

Now, because the operators ${\mathcal L}$, ${\mathcal M}$, ${\mathcal N}$, ${\mathcal O}$, ${\mathcal P}$, and ${\mathcal Q}$ defined in (\ref{def-i}) to (\ref{def-f}) will not commute with each other in general if their constituent operators ${\mathcal R}$, ${\mathcal S}$, ${\mathcal T}$, and ${\mathcal U}$ are non-commuting, the state ${|\,\xi\,\rangle}$ will not be an eigenstate of the operator ${\mathcal Y}$ defined in (\ref{line}). Moreover, while a dispersion-free state $|\,\psi,\,\lambda)$ would pick out one of the eigenvalues ${\mathscr Y}$ of ${\mathcal Y}$, it will not be equal to the linear sum of the corresponding eigenvalues ${\mathscr L}$, ${\mathscr M}$, ${\mathscr N}$, ${\mathscr O}$, ${\mathscr P}$, and ${\mathscr Q}$ in general,
\begin{equation}
{\mathscr Y}\not=rs\,{\mathscr L} + rt\,{\mathscr M} + st\,{\mathscr N} + ru\,{\mathscr O} + tu\,{\mathscr P} + su\,{\mathscr Q}\,,
\end{equation}
even if we assume that the operators ${\mathcal X}$ and ${\mathcal Y}$ commute with each other so that $(\,\psi,\,\lambda\,|\,{\mathcal Y}\,|\,\psi,\,\lambda\,)={\mathscr Y}$ is an eigenvalue of ${\mathcal Y}$. That is to say, just like the eigenvalue ${\mathscr X}$ of ${\mathcal X}$, the eigenvalue ${\mathscr Y}$ of ${\mathcal Y}$ is also a nonlinear function in general. On the other hand, because we wish to prove that the eigenvalue of the sum ${r\,{\mathcal R} + s\,{\mathcal S} + t\,{\mathcal T} + u\,{\mathcal U}}$ of the operators ${\mathcal R}$, ${\mathcal S}$, ${\mathcal T}$, and ${\mathcal U}$ is not equal to the sum $r\,{\mathscr R} + s\,{\mathscr S} + t\,{\mathscr T} + u\,{\mathscr U}$ of the individual eigenvalues of the operators ${\mathcal R}$, ${\mathcal S}$, ${\mathcal T}$, and ${\mathcal U}$ unless they commute with each other, we must make sure that the eigenvalue ${\mathscr Y}$ does not vanish for the unlikely case in\break which the operators ${\mathcal L}$, ${\mathcal M}$, ${\mathcal N}$, ${\mathcal O}$, ${\mathcal P}$, and ${\mathcal Q}$ commute with each other. But even in that unlikely case, we would have
\begin{equation}
\overline{\mathscr Y}=rs\,{\mathscr L} + rt\,{\mathscr M} + st\,{\mathscr N} + ru\,{\mathscr O} + tu\,{\mathscr P} + su\,{\mathscr Q}\,
\end{equation}
as eigenvalue of the operator ${\mathcal Y}$ defined in (\ref{line}), and consequently the eigenvalue ${\mathscr X}$ in (\ref{A12}) will at best reduce to
\begin{equation}
{\mathscr X}=\pm\sqrt{\big\{r\,{\mathscr R} + s\,{\mathscr S} + t\,{\mathscr T} + u\,{\mathscr U}\big\}^2 + rs\,{\mathscr L} + rt\,{\mathscr M} + st\,{\mathscr N} + ru\,{\mathscr O} + tu\,{\mathscr P} + su\,{\mathscr Q}\,}.
\end{equation}
In other words, even in such an unlikely case ${\mathscr Y}$ will not vanish, and consequently the eigenvalue ${\mathscr X}$ will not reduce$\;$to
\begin{equation}
\overline{\mathscr X}=r\,{\mathscr R} + s\,{\mathscr S} + t\,{\mathscr T} + u\,{\mathscr U}. \label{A18}
\end{equation}
Consequently, unless $\left(\,\psi,\,\lambda\,|\,{\mathcal Y}(c)\,|\,\psi,\,\lambda\,\right)\equiv0$, the expectation value of ${\cal X}(c)$ equating the average of ${\mathscr X}(c,\,\lambda)$ will be
\begin{align}
\langle\,\psi\,|\,{\mathcal X}(c)\,|\,\psi\,\rangle &=
\!\int_{\mathscr L}{\mathscr X}(c,\,{\lambda})\;p(\lambda)\,d\lambda \\
&=\!\int_{\mathscr L}\!\left[\pm\sqrt{\big\{r\,{\mathscr R}(c,\,{\lambda}) + s\,{\mathscr S}(c,\,{\lambda}) + t\,{\mathscr T}(c,\,{\lambda}) + u\,{\mathscr U}(c,\,{\lambda})\big\}^2 + \,\left(\,\psi,\,\lambda\,|\,{\mathcal Y}(c)\,|\,\psi,\,\lambda\,\right)\,}\,\right]p(\lambda)\,d\lambda \label{A19} \\
&\not=\!\int_{\mathscr L}\!\left[\pm\sqrt{\big\{r\,{\mathscr R}(c,\,{\lambda}) + s\,{\mathscr S}(c,\,{\lambda}) + t\,{\mathscr T}(c,\,{\lambda}) + u\,{\mathscr U}(c,\,{\lambda})\big\}^2 + \,\overline{\mathscr Y}(c,\,\lambda)}\,\right]p(\lambda)\,d\lambda\;\;\text{if}\;[\,{\mathcal X},\,{\mathcal Y}\,]\not=0\label{A20} \\
&\not=\!\int_{\mathscr L}
\pm\big\{r\,{\mathscr R}(c,\,{\lambda}) + s\,{\mathscr S}(c,\,{\lambda}) + t\,{\mathscr T}(c,\,{\lambda}) + u\,{\mathscr U}(c,\,{\lambda})\big\}\;p(\lambda)\,d\lambda\;\;\text{if}\;{\mathscr L},\,{\mathscr M},\,{\mathscr N},\,{\mathscr O},\,{\mathscr P},\,{\mathscr Q}\not=0, \label{A21}
\end{align}
where $c$ indicates the contexts of experiments as discussed in Section~\ref{Sec-B}. The above result confirms the inequality (\ref{incorrect}) we discussed in Section~\ref{Sec-F}. Note that, because ${\mathscr X}(c,\,\lambda)$ and ${\mathscr Y}(c,\,\lambda)$ are highly {\it nonlinear} functions in general (recall, {\it e.g.}, that $\sqrt{x^2\pm y^2\,}\not=\sqrt{x^2}\,\pm\sqrt{y^2}\,$), the inequality in (\ref{A21}) can reduce to equality {\it if and only if} the operators ${\mathcal R}$, ${\mathcal S}$, ${\mathcal T}$, and ${\mathcal U}$ commute with each other. In that case, the operators ${\mathcal L}$, ${\mathcal M}$, ${\mathcal N}$, ${\mathcal O}$, ${\mathcal P}$, and ${\mathcal Q}$ defined in (\ref{def-i}) to (\ref{def-f}) will also commute with each other, as well as being null operators, with each of the eigenvalues ${\mathscr L}$, ${\mathscr M}$, ${\mathscr N}$, ${\mathscr O}$, ${\mathscr P}$, and ${\mathscr Q}$\break reducing to zero. Consequently, in that case $\left(\,\psi,\,\lambda\,|\,{\mathcal Y}(c)\,|\,\psi,\,\lambda\,\right)$ will vanish identically and (\ref{A12}) will reduce to (\ref{A18}). 

It is now straightforward to deduce the operator ${\widetilde{\Theta}}({\bf a},{\bf a'},{\bf b},{\bf b'})$ specified in (\ref{theta}) using (\ref{nop}). For this purpose, we first note that for the Bell-CHSH sum (\ref{combi}) the real numbers ${r=s=t=+1}$ and ${u=-1}$, and therefore (\ref{A18}) simplifies to
\begin{equation}
\overline{\mathscr X}(c,\,\lambda)={\mathscr R}(c,\,\lambda) + {\mathscr S}(c,\,\lambda) + {\mathscr T}(c,\,\lambda) - {\mathscr U}(c,\,\lambda). \label{Bell-e}
\end{equation}
This quantity is tacitly assumed in the derivation of Bell's theorem to be the eigenvalue of the summed operator (\ref{op}), implying the following identifications:
\begin{align}
{\mathscr A}({\bf a},\,{\lambda})\,{\mathscr B}({\bf b},\,{\lambda})&\equiv {\mathscr R}({\bf a},\,{\bf b},\,{\lambda}) \notag \\
&=\pm1\;\;\text{is an eigenvalue of the observable}\;\;{\mathcal R}({\bf a},\,{\bf b})\equiv{\boldsymbol\sigma}_1\cdot{\bf a}\,\otimes\,{\boldsymbol\sigma}_2\cdot{\bf b}\,, \label{ab} \\
{\mathscr A}({\bf a},\,{\lambda})\,{\mathscr B}({\bf b'},\,{\lambda})&\equiv {\mathscr S}({\bf a},\,{\bf b'},\,{\lambda}) \notag \\
&=\pm1\;\;\text{is an eigenvalue of the observable}\;\;{\mathcal S}({\bf a},\,{\bf b'})\equiv{\boldsymbol\sigma}_1\cdot{\bf a}\,\otimes\,{\boldsymbol\sigma}_2\cdot{\bf b'}\,, \\
{\mathscr A}({\bf a'},\,{\lambda})\,{\mathscr B}({\bf b},\,{\lambda})&\equiv {\mathscr T}({\bf a'},\,{\bf b},\,{\lambda}) \notag \\
&=\pm1\;\;\text{is an eigenvalue of the observable}\;\;{\mathcal T}({\bf a'},\,{\bf b})\equiv{\boldsymbol\sigma}_1\cdot{\bf a'}\,\otimes\,{\boldsymbol\sigma}_2\cdot{\bf b}\,, \\
\text{and}\;\;\;{\mathscr A}({\bf a'},\,{\lambda})\,{\mathscr B}({\bf b'},\,{\lambda})&\equiv {\mathscr U}({\bf a'},\,{\bf b'},\,{\lambda}) \notag \\
&=\pm1\;\;\text{is an eigenvalue of the observable}\;\;{\mathcal U}({\bf a'},\,{\bf b'})\equiv{\boldsymbol\sigma}_1\cdot{\bf a'}\,\otimes\,{\boldsymbol\sigma}_2\cdot{\bf b'}. \label{a'b'}
\end{align}
The non-commuting part of the operator (\ref{op}) can therefore be identified using (\ref{line}) and the above identifications~as
\begin{equation}
{\widetilde{\Theta}}({\bf a},{\bf a'},{\bf b},{\bf b'}) = \big\{{\mathcal L} + {\mathcal M} + {\mathcal N} - {\mathcal O} - {\mathcal P} - {\mathcal Q}\big\}({\bf a},{\bf a'},{\bf b},{\bf b'})\,, \label{notline}
\end{equation}
where the operators ${\mathcal L}$, ${\mathcal M}$, ${\mathcal N}$, ${\mathcal O}$, ${\mathcal P}$, and ${\mathcal Q}$ are defined in (\ref{def-i}) to (\ref{def-f}). The result is the operator specified in (\ref{theta}). 

\section{Establishing bounds on the magnitude of the vector ${\mathbf n}$ defined in (\ref{vec})} \label{B}

The vector ${\bf n}$ defined in (\ref{vec}) is a function of four unit vectors, ${\bf a}$, ${\bf a'}$, ${\bf b}$, and ${\bf b'}$, in ${\mathrm{I\!R}^3}$, and involves various cross products among these vectors. Consequently, as the vectors ${\bf a}$, ${\bf a'}$, ${\bf b}$, and ${\bf b'}$ vary in their directions within ${\mathrm{I\!R}^3}$ due to various choices made by Alice and Bob, the extremum values of the magnitude $||{\bf n}||$ are obtained by setting the vectors\break orthogonal to each other, with angles between them set to 90 or 270 degrees. However, in three dimensions that is possible only for three of the four vectors, so one of the four would have to be set either parallel or anti-parallel to one of the remaining three. Therefore, let us first choose to set ${\bf b'}=-{\bf b}$. Substituting this into (\ref{vec}) then gives ${\bf n}={\bf 0}$, and\break thus $||{\bf n}||=0$.~We have thus found the lower bound on the magnitude $||{\bf n}||$.~To determine the upper bound on $||{\bf n}||$, we set ${\bf a'}=-{\bf a}$ instead. Substituting this into (\ref{vec}) reduces the vector ${\bf n}$ to the following function of ${\bf a}$, ${\bf a'}$, ${\bf b}$ and~${\bf b'}$:
\begin{equation}
{\bf n} = 2\,\big\{\left({\bf a}\times{\bf b'}\right)\times\left({\bf a}\times{\bf b}\right)\big\}.
\end{equation}
Consequently, in this case, the magnitude of the vector ${\bf n}$ works out to be
\begin{align}
||{\bf n}|| &= 2\,||({\bf a}\times{\bf b'})||\,||({\bf a}\times{\bf b})||\,\sin\beta_{({\bf a}\times{\bf b'}),({\bf a}\times{\bf b})} \\
&=2\,\big\{||{\bf a}||\,||{\bf b'}||\,\sin\beta_{{\bf a},{\bf b'}}\big\}\,\big\{||{\bf a}||\,||{\bf b}||\,\sin\beta_{{\bf a},{\bf b}}\big\}\,\big\{\sin\beta_{({\bf a}\times{\bf b'}),({\bf a}\times{\bf b})}\big\},
\end{align}
where $\beta_{{\bf a},{\bf b}}$ is the angle between ${\bf a}$ and ${\bf b}$, {\it etc}. But since the vectors ${\bf a}$, ${\bf a'}$, ${\bf b}$, and ${\bf b'}$ are all unit vectors and we have set them orthogonal to each other (apart from ${\bf a'}=-{\bf a}$), we obtain $||{\bf n}|| = 2$ as the maximum possible value for the magnitude of ${\bf n}$. We have thus established the following bounds on the magnitude of the vector ${\bf n}$ as specified in (\ref{vec}):
\begin{equation}
0\,\leqslant\,||{\bf n}||\,\leqslant\,2.
\end{equation}

\section{Consistency check for the critique of Bell's theorem presented in Section~\ref{Sec-F}}

With the correct eigenvalue (\ref{correct}) of the summed operator $\widetilde{\Omega}(\tilde{c})$ worked out in the Eq.~(\ref{op}) of Section~\ref{Sec-F}, namely,
\begin{equation}
{\widetilde{\omega}}\!=\pm\sqrt{\big\{{\mathscr A}({\bf a},\,\lambda)\,{\mathscr B}({\bf b},\,\lambda)+{\mathscr A}({\bf a},\,\lambda)\,{\mathscr B}({\bf b'},\,\lambda)+{\mathscr A}({\bf a'},\,\lambda)\,{\mathscr B}({\bf b},\,\lambda)-{\mathscr A}({\bf a'},\,\lambda)\,{\mathscr B}({\bf b'},\,\lambda) \big\}^2 + (\Psi,\,\lambda\,|\,{\widetilde{\Theta}}\,|\,\Psi,\,\lambda)\,}\;, \label{correct-2}
\end{equation}
the RHS of (\ref{corbon}) is its expectation value involving functions of the form $X(\lambda)$ and $Y(\lambda)$ with probability density $p(\lambda)$:
\begin{equation}
\int_{\mathscr L} {\widetilde{\omega}}({\bf a},{\bf a'},{\bf b},{\bf b'},\lambda)\;p(\lambda)\,d\lambda \;=:\; {\cal{E}}\left[\pm\sqrt{X^2(\lambda)+Y(\lambda)}\;\right] =\int_{\!\mathscr{L}} \pm\sqrt{X^2(\lambda) + Y(\lambda)} \;\;p(\lambda)\,d\lambda\,. \label{E2}
\end{equation}
But the expectation value of the function $Y(\lambda):=(\Psi,\,\lambda\,|\,{\widetilde{\Theta}}\,|\,\Psi,\,\lambda)=(\Psi,\,\lambda\,|\,2\,{\boldsymbol\sigma}\cdot{\bf n}\,|\,\Psi,\,\lambda)=\pm2\,||{\bf n}||$ may vanish $\forall\,{\bf n}$:
\begin{equation}
{\cal{E}}\big[Y(\lambda)\big] = \int_{\!\mathscr{L}} Y(\lambda) \,\;p(\lambda)\,d\lambda\;=\int_{\!\mathscr{L}} (\Psi,\,\lambda\,|\,2\,{\boldsymbol\sigma}\cdot{\bf n}\,|\,\Psi,\,\lambda)\;p(\lambda)\,d\lambda = \left(+\,2\,||{\bf n}||\times\frac{1}{2}\right) + \left(-\,2\,||{\bf n}||\times\frac{1}{2}\right) = 0\,. \label{E3} 
\end{equation}
We can also see this using the rotational invariance of the singlet state (\ref{single}) (cf. \cite{Christian}), and the hypothesis (\ref{77}) in reverse: 
\begin{equation}
{\cal{E}}\big[Y(\lambda)\big] = \int_{\!\mathscr{L}} Y(\lambda) \,\;p(\lambda)\,d\lambda\;=\int_{\!\mathscr{L}} (\Psi,\,\lambda\,|\,2\,{\boldsymbol\sigma}\cdot{\bf n}\,|\,\Psi,\,\lambda)\;p(\lambda)\,d\lambda =2\times\langle\Psi|\,{\boldsymbol\sigma}\cdot{\bf n}\,|\Psi\rangle = \,0\,,\;\forall\;\,{\bf n}. \label{E44}  
\end{equation}
As a result, suspicion may arise that the eigenvalue (\ref{correct-2}) may reduce to the usual Bell-CHSH form (\ref{bell}) for ``large $N$":  
\begin{equation}
{\cal{E}}\left[\pm\sqrt{X^2(\lambda)+Y(\lambda)}\;\right] \xrightarrow{\;\;``\text{large}\; N"\;\;}\,
{\cal{E}}\big[X(\lambda)\big] =\!\!\int_{\!\mathscr{L}} X(\lambda) \,\;p(\lambda)\,d\lambda\,. \label{E4}
\end{equation}
If so, then that would vindicate Bell's mistaken assumption of linear additivity of expectation values in (23), {\it even for non-commuting observables}. Indeed, as we discussed in Section~\ref{Sec-E}, one popular defense of Bell's theorem relies on a {\it statistical} justification for the stringent bounds of $\pm2$ on the Bell-CHSH correlator derived in (\ref{chsh}), in ``large $N$" limit. 

To recognize that such a hope to save Bell's theorem on statistical grounds is misplaced, let us isolate the function \(|X(\lambda)|\) from the expression $\sqrt{X^2(\lambda)+Y(\lambda)}\,$ by making use of the following easily verifiable, purely algebraic identity:
\begin{align}
\sqrt{X^2(\lambda)+Y(\lambda)} \;\equiv\; |X(\lambda)| + \frac{Y(\lambda)}{|X(\lambda)|+\sqrt{X^2(\lambda)+Y(\lambda)}\,} \,=:\, |X(\lambda)| + Y(\lambda)\!\times\! f[X(\lambda),Y(\lambda)]\,, \label{E5}
\end{align}where it is convenient to define the functional $f[X(\lambda),Y(\lambda)]$ as
\begin{equation}
f[X(\lambda),Y(\lambda)] \,:=\, \frac{1}{|X(\lambda)|+\sqrt{X^2(\lambda)+Y(\lambda)}\,}\,. \label{E6}
\end{equation}
Using (\ref{E5}) and the additivity of expectation values, we can now express the expectation function in (\ref{E2}) equivalently~as
\begin{align}
{\cal{E}}\left[\pm\sqrt{X^2(\lambda)+Y(\lambda)}\,\right] \,\equiv\; \pm\left\{\,{\cal{E}}[\,|X(\lambda)|\,] + \, {\cal{E}}[\,Y(\lambda)\!\times\!f[X(\lambda),Y(\lambda)]\,]\right\}.
\end{align}
However, the function ${\cal{E}}[\,Y(\lambda)\!\times\!f[X(\lambda),Y(\lambda)]\,]$ cannot be factorized into a product of ${\cal{E}}[\,Y(\lambda)\,]$ and ${\cal{E}}[\,f[X(\lambda),Y(\lambda)]\,]$:
\begin{align}
{\cal{E}}[\,Y(\lambda)\!\times\!f[X(\lambda),Y(\lambda)]\,] \not=\,{\cal{E}}[\,Y(\lambda)\,]\times{\cal{E}}[\,f[X(\lambda),Y(\lambda)]\,]\,.
\end{align}
This is because $f[X(\lambda), Y(\lambda)]$ is a functional of {\it both}$\,$ random variables $X(\lambda)$ and $Y(\lambda)$ appearing in (\ref{E2}). Consequently, 
\begin{align}
{\cal{E}}\left[\pm\sqrt{X^2(\lambda)+Y(\lambda)\,}\,\right]
\not=\pm\left\{\,{\cal{E}}[\,|X(\lambda)|\,] +\, {\cal{E}}[\,Y(\lambda)\,]\times{\cal{E}}[\,f[X(\lambda),Y(\lambda)]\,]\right\}.
\end{align}
If it were possible to write the above inequality as an equality, then, in the ``large $N$'' limit, we could use (\ref{E3}) or (\ref{E44}) to infer ${\cal{E}}[\,Y(\lambda)\,]=0$ even for the non-commuting observables, thereby reducing ${\cal{E}}\left[\pm\sqrt{X^2(\lambda)+Y(\lambda)\,}\,\right]$ to $\,{\cal{E}}[\,X(\lambda)\,]$, as shown in (\ref{E4}).~But the vanishing expectation of $Y(\lambda)$, {\it i.e.}, ${\cal{E}}[\,Y(\lambda)\,]=0$, says nothing about ${\cal{E}}[\,Y(\lambda)\times f[X(\lambda),Y(\lambda)]\,]$, {\it i.e.}, the expectation of the product $Y(\lambda)\times f[X(\lambda),Y(\lambda)]$, unless the functional $f[X(\lambda),Y(\lambda)]$ were independent of (or uncorrelated with) the function $Y(\lambda)$. If the latter were the case, then we could indeed factorize ${\cal{E}}[\,Y(\lambda)\times f[X(\lambda)]\,]$ to ${\cal{E}}[\,Y(\lambda)\,]\times\,{\cal{E}}[\,f[X(\lambda)]\,]$, and then reduce ${\cal{E}}\left[\pm\sqrt{X^2(\lambda)+Y(\lambda)\,}\,\right]$ to $\,{\cal{E}}[\,X(\lambda)\,]$, as in (\ref{E4}). But that is clearly not the case, as we see from (\ref{E5}) and (\ref{E6}). This eliminates the stated suspicion and upholds our critique of Bell's theorem.

\def\bibsection{\section*{\refname}}

\end{document}